\documentclass[twocolumn]{aastex62}

\usepackage{xcolor}

\received{\today}
\revised{}
\accepted{}
\submitjournal{ApJ}

\shorttitle{III.  Insights from SPH simulations with chemo-photometric implementation.}
\shortauthors{Mazzei et al.}

\begin{document}

\title{Investigating early-type galaxy evolution with a multi-wavelength approach\\
 III.  Insights from SPH simulations with chemo-photometric implementation.}

\correspondingauthor{Paola Mazzei}
\email{paola.mazzei@inaf.it}

\author[0000-0002-8004-1034]{Paola Mazzei}
\affil{INAF Osservatorio Astronomico Padova, Vicolo dell'Osservatorio 5, 35122, Padova, Italy}

\author{Roberto Rampazzo}
\affiliation{INAF Osservatorio Astrofisico di Asiago,
Via dell'Osservatorio, 8, 36012, Vicenza, Italy}

\author{Antonietta Marino}
\affil{INAF Osservatorio Astronomico Padova, Vicolo dell'Osservatorio 5, 35122, Padova, Italy}

\author{Ginevra Trinchieri}
\affiliation{INAF Osservatorio Astronomico di Brera, Via Brera 28, 20121 Milano, Italy}

\author{Michela Uslenghi}
\affiliation{INAF-IASF, via E. Bassini 15, 20133 Milano, Italy}

\author{Anna Wolter}
\affiliation{INAF Osservatorio Astronomico di Brera, Via Brera 28, 20121 Milano, Italy}

\begin{abstract}

 We are exploring  galaxy evolution in low density environments exploiting smooth particle hydrodynamic  simulations including chemo-photometric implementation. From a large grid of simulations of galaxy encounters and mergers  starting from triaxial halos of gas e dark matter,  we single out the simulations matching the global properties of our targets. These simulations  are used to give insights into their evolution.  We focus on 11 early-type galaxies selected  because of their nearly passive stage of evolution in the nuclear region. However, a variety of UV features are detected in more than half of these galaxies.   We find no significant  differences in the formation mechanisms between galaxies with or without UV features. Major and minor mergers are able to reproduce their  peculiar UV morphologies, galaxy encounters are more suitable for 'normal' early-type galaxies. Their star formation rate self-quenches several Gyr later the merger/encounter occurred, via gas exhaustion and stellar feedback, moving the galaxy from blue to red colors, driving the galaxy  transformation. The length of the quenching is mass dependent and lasts from 1 to 5 Gyr or more in the less massive systems.  All our targets are gas rich  at redshift 1.  Three of them assembled at most 40\% of their current stellar mass at z$>$1, and seven  assembled more than 50\%   between redshift 0.5 and 1. Their stellar mass grows with 4\% by crossing the Green Valley before  reaching  their current position on the NUV$-$r vs. M$_r$   diagram.

\end{abstract}


\section{Introduction} \label{Intro}
Our understanding of early-type galaxies (Es+S0s, ETGs hereafter) evolution has been greatly improved in the recent years
thanks to multi-$\lambda$ observations from satellites and ground-based telescopes.
Relatively large amount of cold residual gas (10$^7$-10$^9$ M$_{\sun}$) is now considered a common property (50\%) of these galaxies 
\citep[e.g.][and references therein]{Young18}.
The combination of the  Ultraviolet (UV) data provided by {\it GALaxy evolution EXplorer}
  \citep[{\it GALEX} hereafter;][]{Martin05} 
with the optical view given by the Sloan Digital Sky Survey  \citep[SDSS\,hereafter;][]{Stoughton02} 
has greatly undermined the standard picture of ETGs as passively evolving systems. 
Statistical studies  \citep[e.g.][]{Kaviraj07,Schawinski07} found that about 30\% of
massive ETGs show recent/ongoing Star Formation (SF), with the largest incidence in galaxies 
located in low density environments. However, also $\sim$27/43\% of cluster ETGs 
show signatures of recent SF  \citep[][respectively]{Yi11,Hernandez14}.
UV bright extended structures, e.g. rings, arm-like features, tails, sometimes completely distinct
in shape from the optical galaxy body \citep[][]{ Rampazzo07, Rampazzo11, Rampazzo2018, Marino09, Marino11a, Marino11b, Jeong09, Thilker10} are found, often associated to HI emission \citep{Salim10,Salim12}. The color magnitude diagram of  galaxies, using SDSS and {\it GALEX} magnitudes,  $(NUV-r)~vs.~M_r$ (CMD hereafter),
shows a strong morphological segregation: Spirals lie in the Blue Cloud (BC) while ETGs are mainly located in the Red Sequence (RS) \citep[e.g.][]{Yi05,Salim07,Wyder07}.
An intermediate area, called Green Valley (GV), exists, irrespective of the
environment.  The CMD is a powerful diagnostic tool to detect very low levels of SF. 

The spectral analysis with mid-infrared (MIR hereafter) {\tt Spitzer}-IRS of a large number of ETGs, within $\approx$2-3~$r_e/8$, highlighted the role of Polycyclic Aromatic Hydrocarbon features (PAHs)  as tracers of episodes of SF occurred between 1 and 2.5 \,Gyr ago,
depending on metallicity  \citep{Bressan06,Kaneda08,Vega10,Panuzzo11,Rampazzo13,Nanni13}. 
This paper, the  third of a series based on  {\tt SWIFT} 
\citep{swift1,swift2,swift3} multi-$\lambda$
 ({\tt XRT + UVOT}) observations, is dedicated to tracing back
the evolution of  eleven nearby ETGs (Table \ref{tab:tab1}) living in low density environments \citep[][their Table 1]{ Rampazzo17}, for which we collected a large, homogeneous,  multi-$\lambda$ data set, both from our previous papers
\citet[][hereafter Paper~I]{Trinchieri15},  \citet[][hereafter Paper II]{Rampazzo17}, and from the literature,  to constrain our simulations. 
Galaxies are selected because of the nearly passive stage of evolution of their nuclear region 
\citep{Rampazzo13}.   Most of these galaxies, which could be considered as templates of {\it nearly-dead} ETGs, show indeed 
a manifold of bright and peculiar structures in {\it  SWIFT}-{\tt UVOT} images in the UV band (Figure 7 of  Paper~I). 
By analyzing  their  UV luminosity profiles, previously unexplored (Paper II), we found  that  this spectral range highlights the presence of a stellar disk.  
Moreover, accretion episodes have characterized the recent evolution of  many of these ETGs. Such episodes left signatures of recent SF either in the nuclei and/or in the galaxy outskirts.  The recent SF points toward  wet accretion episodes, since dissipation is the source
of the underlying disk structure highlighted by  the analysis (Paper~II).

Morphology, colors, and the star formation rate (SFR) primarily depend on small-scale ($<$1 Mpc) environment \citep{Hogg04, Kauffmann04, Wetzel12}.
This result has been extended by studying galaxy group samples: they show that colors and SF history most directly depend on the properties of the host dark matter (DM) halo  \citep{BB07, Wei10, Tinker12}, in agreement with results of smooth particle hydrodynamical (SPH) simulations by \citet[][MC03, hereafter]{MC03}. In this context, the investigation of the evolution of group members in the nearby universe acquires a great cosmological interest because more than half of galaxies reside in such environments. Furthermore, since the velocity dispersion of galaxies is significantly lower in groups than in clusters, the merger probability and the effects of interaction on galaxy evolution are much higher. Consequently, groups provide a zoom-in on phenomena driving the galaxy  evolution before galaxies fall into denser environments \citep[e.g.,][]{Wilman09, Just10}. 
Starting from the pioneering works of \citet{TT72} and reference therein, several papers  contributed to shed light on the important role of merges/interactions in galaxy evolution,  \citet{Toomre77}, \citet{Combes90}, \citet{BeH96}, \citet{MeH94, MeH96}, \citet{NB03}, \citet{BeC05}, \citet{DM07} to name a few,  up to the most recent papers of  \citet{EM18} and \citet[][and references therein]{Martin18}.
Dissipative merger simulations of  \citet{EM18} start from systems just formed,  composed of a spherical non-rotating  DM halo, and by a disk of gas particles with or without the presence of a stellar bulge.  These simulations explore about 3-3.5\,Gyr of evolution.
\citet{Martin18} focus on processes triggering galaxy transformations of massive galaxies (M$>$ 10$^{10}$\,M$_{\odot}$) exploiting  {\sl Horizon-AGN} cosmological hydrodynamic simulations  by \citet{Kaviraj17}. These simulations, based on an adaptive mesh refinement code ({\tt RAMSES}) and including the baryon treatment with stellar and AGN feedback, are able to resolve baryonic physics on kpc scale,  larger than we use in this paper (Sect. 3.)
They derived important statistical assessments about the processes that drive morphological transformation across cosmic time.
 
We explore the merger/interaction scenario focusing on low density environments.  Our simulations start from collapsing triaxial systems composed of DM and gas and combine Smooth Particle Hydrodynamic (SPH) code with  Chemo-Photometric Implementation based on  Evolutionary Population Synthesis (EPS) models (SPH-CPI simulations hereafter)  providing the spectral energy distribution (SED) at each snapshot.
From a large grid of SPH-CPI simulations, we select for each galaxy the one that matches the current, global properties of our target, in particular i) the SED  extended over four orders of magnitude in $\lambda$, ii) the absolute B-band magnitude and iii) the optical and UV morphologies, as confirmed by iv) the luminosity profiles. Moreover, the selected simulation must account for v) the $HI$ gas mass, vi) the hot gas X-ray luminosity, and vii) the available kinematic data.
These simulations, anchored to the local properties of our targets, are used to give insights into galaxy evolution and, in particular, to shed light on their quenching phase which is identified by their rest-frame CMDs.  Therefore, we study the  galaxy transformation by looking at the behaviour of
 SFR, gas accretion history,  total mass growth of different components (baryons,  stars, DM), and, in particular, the CMD diagram.  This is a useful tool to
capture very low level of residual SFR,  highlighting the length of different phases leading to quenching and galaxy transformation from the Blue Cloud to the Red Sequence.

The SED accounts for chemical evolution, stellar emission, internal extinction, and re-emission by dust in a self-consistent way, as described in previous works \citep[][and references therein]{Spa09, Spa12}. This extends more than four orders of magnitude in wavelength, from 0.05 to 1000\,$\mu$m. Each simulation in the grid self-consistently provides morphological, dynamic, and chemo-photometric evolution. 
We applied already the same approach to ETGs belonging to the groups USGC~367 and LGG~225 \citep{Mazzei14a}, and to two S0 galaxies, namely NGC~3626 and NGC~1533, in \citet{Mazzei14b}, where we match 
photometric, structural (e.g. disk vs. bulge) and kinematical (gas vs. stars) properties, showing that a major merger  (1:1)  accounts  for the  structural and photometric  transformations expected in these systems \citep{Querejeta15}.  Furtheremore, our approach allowed us to fit the current global properties of i) several other ETGs \citep{Spa09,Spa12, Tri12, Bettoni11, Bettoni12, Bettoni14},   ii)  the blue dwarf galaxy  UGC~7639 \citep{Buson2015},  iii)  a few late-type galaxies of different luminosity classes \citep{Bettoni11, Bettoni14, Mazzei18},  iv) a  pre-merger case, NGC 454 \citep{Plana17}, and v) the false pair NGC~3447/3447A \citep{Mazzei18},
showing that this a powerful tool to investigate galaxy  evolution.

The plan of the paper  is the following. In Section~\ref{Sample} we review the  
galaxy sample and in Section \ref{Recipes} we summarize the  recipes of our SPH-CPI simulations.
We obtain the  evolution of these galaxies anchoring  the simulations to match the  global current properties  of  each ETG.
 Results are presented in 
Section \ref{Results}, while the details of the fit of each target are listed in the {Appendix}.
Discussion and Conclusion follow  in Section~\ref{Discussion} and  Section~\ref{Summary} respectively.
 
Hereafter we use the following cosmological parameters: H$_0$=67\,km s$^{-1}$\,Mpc$^{-1}$, $\Omega_{\Lambda}$=0.68, $\Omega_{matter}=$0.32 
 \citep{PC14, Calab17} which correspond to a $\Lambda$ cold dark matter ($\Lambda$CDM) model with  Universe  age of 13.81\,Gyr  \citep[][their Table 1]{Calab17}, and
a light travel time (look-back time) of 7.98\,Gyr at z=1 and 5.23\,Gyr at z=0.5.

\begin{deluxetable*}{ccccCrlc}[hbt!]
\tablecaption{The sample of ETGs \label{tab:tab1}}
\tablehead{
\colhead{Galaxy} & \colhead{D$_{25}$} & \colhead{D}  & \colhead{scale} & \colhead{m-M} & \colhead{M$_{B}$} & \colhead{M$_{HI}$} & \colhead{L$_{X}$(gas)}\\
\colhead{} & \colhead{(arcmin)} & \colhead{(Mpc)}  & \colhead{(kpc\,arcmin$^{-1}$)} & \colhead{(mag)}  & \colhead{(mag)} &\colhead{(10$^9$\,M$_{\odot}$)} & \colhead{(10$^{40}$\,erg\,s$^{-1}$)}
}
\startdata
   NGC~1366 &2.1 & 21.1$\pm$2.1 & 6.1&  31.62$\pm$0.50 & -18.89$\pm$0.54 & $<$1.0 & $<$0.03\\
   NGC~1415 &3.7 & 22.7$\pm$2.5 & 6.5&  31.78$\pm$0.55 & -19.23$\pm$0.55 & 1.2\,$^1$ & 0.1    \\
   NGC~1426 &2.9 & 24.1$\pm$2.4  & 7.0  & 31.91$\pm$0.50 & -19.71$\pm$0.52 & ....   & $<$0.03 \\
   NGC~1533 &3.2& 21.4$\pm$2.1 & 6.2  &  31.65$\pm$0.50 & -19.93$\pm$0.52   &  7.4\,$^2$   & $<$0.11\\
   NGC~1543 &3.6& 20.0$\pm$2.0 & 5.8 & 31.50$\pm$0.50 & -20.13$\pm$0.53  & 0.8   &   $<$0.16 \\
   NGC~2685 &4.4& 16.0$\pm$1.6  & 4.8 &  31.02$\pm$0.50& -19.13$\pm$0.51  &  3.0\,$^3$  & $<$0.04\\
   NGC~2974 &3.5& 21.5$\pm$2.0  & 6.2 &  31.66$\pm$0.46&-20.01$\pm$0.60 & 0.7\,$^4$   &  0.2  \\
   NGC~3818 &2.4& 36.3$\pm$3.6 & 10.4  &  32.80$\pm$0.50 & -20.26$\pm$0.58  & ...  & 0.55\\
   NGC~3962 &4.2& 35.3$\pm$3.5  & 10.2 & 32.74$\pm$0.50 &   -21.32$\pm$0.53 & 2.8\,$^5$   & 0.33\\ 
   NGC~7192 & 2.4&37.8$\pm$3.8  & 10.7 &  32.89$\pm$0.50 & -20.84$\pm$0.51   & 0.7\,$^5$ & 1.0\\
   IC~2006  &2.3 &20.2$\pm$2.0 & 5.9 &  31.53$\pm$0.50 & -19.34$\pm$0.51 & 0.3 & 0.08\\   
 \enddata
\tablecomments{The apparent diameters (col. 2)  and the  adopted distances (col. 3) are derived from 
the Extragalactic Distance Database (EDD: http://edd.ifa.hawaii.edu), as in Papers I and II.
Absolute total magnitudes in col. 6 are derived from col. 5 using B-band observed total magnitudes and
extinction corrections from {\tt Hyperleda}  catalogue \citep[http://leda.univ-lyon1.fr][]{Makarov2014}.
The HI masses (col. 7) are obtained using  distances in col. 3 and  1.4\,GHz fluxes from {\tt NED{\footnote{
The NASA/IPAC Extragalactic Database (NED) is operated by the Jet Propulsion Laboratory, California Institute of Technology, under contract with the National Aeronautics and Space Administration
}}}: 
 $^1$  \citet{Courtois15}; $^2$ \citet{Ryan-Weber2003}; $^3$\citet{Jozsa09}; $^4$ \citet{Kim88};
$^5$ \citet{Serra10}.
X-ray gas luminosities (col. 8) are from Table ~7 of Paper I.
}
\end{deluxetable*}
\section{The sample}
\label{Sample}

The sample is presented in
Table~\ref{tab:tab1}, where all the main observed characteristics are listed for each galaxy. All galaxies are classified as ETGs (see Paper~II, Table 1).  A transition case between early 
and later types is NGC~1415 whose classification has, however, 
a large uncertainty (Paper II). 
Their absolute M$_{B}$  magnitudes,  from  -18.9\,mag to -21.3\,mag, are all extinction corrected.
These ETGs  are gas rich on average,
with HI masses of the order of 10$^9$ M$_\odot$.
Table~1 of  Paper II  collects  their environmental properties which span a large range of local densities:   $0.13 \leq \rho_{xyz}$ (gal Mpc$^{-3}$) $\leq 0.95$.
We select here only galaxies with the nuclear region, 2-3$\times$ r$_e$/8,
in a nearly passive stage of evolution (col. 4 in Table 1  of  Paper~I).
Many of these galaxies,  which could be considered  as templates of {\it nearly-dead} ETGs, nonetheless  show a manifold
of bright and peculiar structures in {\it  SWIFT}-{\tt UVOT} images in the UV band.
Their  UV and optical images have been presented for each band in Figure~7 of  Paper~I. Their three colors composite images, both in the UV and in the optical bands,
are show in Figure 4 to Figure 14 of Paper II. We showed in that paper that a single S\'ersic law \citep{Sersic68} is a good fit to their surface brightness luminosity profiles with decreasing values of the  S\'ersic index, n, for decreasing wavelength.  The  S\'ersic law is widely used for ETGs  since it is a generalization of the r$^{1/4}$  \citet{deV48} law \citep[see, e.g.,][]{Caon93}. Special cases are n$=$1, the value for an exponential profile, and n$=$0.5 for
a Gaussian luminosity profile. Galaxies with n  higher than 1 have a steep luminosity profile in their nuclear regions and extended outskirts. Values lower than 1 indicate a flat nuclear region and more sharply truncated outskirts (see also Paper II).  The values of their UV  S\'ersic indices highlighted the presence of  underlying  disks and  the role of  dissipative mechanisms.

According to their UV morphology, from PaperII, the sample can be divided into two  sets. 
Five ETGs, namely NGC~1366, NGC~1426, NGC~3818, NGC~3962, and NGC~7192
do not show any remarkable feature both in the optical and in the UV bands. 
The remaining six ETGs, namely NGC~1415, NGC~1533, NGC~1543, 
NGC~2685, NGC~2974 and NGC  IC~2006, show bright, peculiar, ring/arm-like UV features.\\

\begin{deluxetable*}{ccccCrlcccc}[hbt!]
\tablecaption{Initial parameters of SPH-CPI simulations reproducing the 11 ETGs
\label{tab2} }
\tablehead{
\colhead{Name} &\colhead{M$_{tot}$}&\colhead{mass ratio} &\colhead{f$_{gas}$}&\colhead{N$_{part}$}  &\colhead{r$_1$}  & \colhead{r$_2$} &\colhead{v$_1$}  &  \colhead{v$_2$}  &\colhead{spins}    &  \colhead{m/e}   \\
\colhead{}&\colhead{(10$^{10}$M$_{\odot}$)}  &  \colhead{}       &\colhead{}  & \colhead{}       &\colhead{(kpc)}     &\colhead{(kpc)}   &  \colhead{(km\,s$^{-1}$)}                &\colhead{(km\,s$^{-1}$)}  & \colhead{} &   \colhead{}
}
\startdata
 NGC~1366& 110 & 10:1  &0.1& 2.2$\times$10$^5$  &85  & 847    & 7     &   70  & $\updownarrow$ &e \\
 NGC~1415 & 600  &5:1&0.02&    1.2$\times$10$^5$ & 206        & 1032    &   26   & 82   & $\parallel$   &     m     \\
 NGC~1426 & 200    &  1:1 &0.1& 8.0$\times$10$^4$    & 327   & 327  & 71 & 71  & $\parallel$  & m    \\ 
 NGC~1533 & 300   & 2:1 &0.1&  6.0$\times$10$^4$  & 155 & 311 & 60 & 119& $\perp$&m\\
 NGC~1543 & 240 &  5:1  &0.1& 9.6$\times$10$^4$ &  78  & 389    &27 & 134 &$\parallel$& m \\
 NGC~2685 & 150 &  2:1&0.1 &6.0$\times$10$^4$  & 311   & 621&  29& 60& $\perp$&m\\
 NGC~2974 &  400 & 1:1&0.1&8.0$\times$10$^4$ & 466   &  466&  73 & 73&$\parallel$ &m\\
 NGC~3818 &  400 & 1:1  &0.1& 6.0$\times$10$^4$    &  233   &  233  & 104 &  104&$\parallel$& m  \\
 NGC~3962 & 550 & 10:1 &0.1 & 1.1$\times$10$^5$   & 177 & 1779 & 10 & 104  &$\parallel$&e \\ 
 NGC~7192 &400  & 1:1  &0.1& 6.0$\times$10$^4$   &466  &  466 &  73 & 73&$\parallel$ &m\\
 IC~2006  & 300   & 2:1 &0.1 &6.0$\times$10$^4$      & 155  &311 & 60& 119 &$\updownarrow$ &m \\
  \enddata
\tablecomments{Our targets are in col.1,  the total mass of the simulation selected in col.2, the halo mass ratio  in col.3, the initial gas fraction in col.4, the initial number of particles in col.5; positions and velocities of the two halos with respect to  the mass center of the system in col.s 6-9,   the halo spin direction ($\parallel$ parallel; $\perp$ perpendicular;  $\updownarrow$   counter-rotating) in  col.10; label in col.11 indicates  if  the simulation gives rise to a merger (m) or an  encounter/flyby (e).
}
\end{deluxetable*}
\section{ SPH-CPI simulations }  
\label{Recipes}

We  have performed a large grid of SPH-CPI simulations of galaxy encounters and mergers starting from triaxial systems initially composed of DM and  gas. 
As described in several previous works \citep{Mazzei14a, Mazzei14b, Mazzei18}, for each galaxy 
we single out  the simulation 
that matches the current, global properties of our target (see Table \ref{tab:tab1}), in particular i) the SED  extended over four orders of magnitude in $\lambda$, ii) the absolute B-band magnitude and iii) the optical and UV morphologies, as confirmed by iv) the luminosity profiles. Moreover, the selected simulation must account for v) the $HI$ (cold) gas mass, vi) the hot gas X-ray luminosity, and vii) the available kinematic data from the HyperLeda catalogue.
The  simulation that contains the snapshot  which best reproduces all  these  observational constraints,  also provides insights on the galaxy evolution.

All the simulations start from collapsing triaxial systems composed of DM and gas in different proportion and different total masses, as in \citet{CM98}, and \citet[][MC03 hereafter]{MC03}.  In particular,  all the simulated  halos have the same initial conditions, that is,  virial ratio (0.1),  average density, to avoid different collapse time, and  spin parameter  equal to 0.06 aligned with the shorter principal axis of the DM halo.
The initial triaxiality ratio of the DM halos  as detached by the Hubble flow in a cold dark matter (CDM) scenario, $\tau$ = (a$^2$ - b$^2$)/(a$^2$ - c$^2$),  is 0.84  \citep{Mazzei14a, Mazzei14b} where a $>$ b$ >$ c.
 This $\tau$ value is different from the fiducial value  adopted by MC03 (0.58)  and motivated to be closer to the initial  condition  of cosmological halos   as in \citet{Warren1992}, 
\citet[][their Table 1]{CMM06}, and \citet{Schneider12}.
The simulations include self-gravity of gas, stars and DM, radiative cooling, hydrodynamical pressure, 
shock heating, viscosity, star formation (SF), feedback from evolving stars and type-II SNs, and chemical enrichment.
The initial mass function (IMF) is of Salpeter type \citep{Salp55} with upper and lower mass limit 100 and 0.01 M$_{\odot}$ respectively. 
All the model parameters had been tuned in  previous papers devoted to analyzing the evolution of isolated collapsing triaxial
halos, initially composed of DM and gas  \citep[and MC03]{CM98, Mazzei03}. In those papers, the role of the initial spin of the halos, their total mass and
gas fraction,  triaxiality ratio, as well as different IMFs, particle resolutions, SF efficiencies, and values of the feedback parameter, were all examined. 
In particular, a slightly higher SFR compared with the other possibilities examined arises from our choice of IMF parameters (see MC03; their Fig.1); this allows for the lowest stellar feedback strength (63\% less than that in the same simulation with a lower mass limit 0.1\,M$_{\odot}$  instead of 0.01\,M$_{\odot}$), and for the expected rotational support when disk galaxies are formed (MC03).
As pointed out by Kroupa (2012), this slope is almost the same as the universal mass function that links the IMF of galaxies
and stars to those of brown dwarfs, planets, and small bodies (meteoroids, asteroids; Binggeli \& Hascher 2007).
The integrated properties of simulated galaxies, stopped at 15 Gyr have been successfully compared with those of local galaxies  (\citet[ e.g their Fig. 17]{CM98}, \citet{Mazzei03}, \citet{Mazzei04}, their Fig. 8). This means  that  their colors, absolute magnitudes, metallicities, and mass to luminosity ratios, all match the results of simulations.
Each simulation self-consistently provides morphological, dynamic, and chemo-photometric evolution, namely the
SED, at each evolutionary time (which we call a 'snapshot').  
The SED accounts for chemical evolution, internal extinction, and re-emission by dust in a  self-consistent way.  This extends more than  four orders of magnitude in wavelength, from 0.05 to  1000\,$\mu$m.
Additionally, we derive the X-ray luminosity of the hot gas (T$\ge$3$\times$10$^6$\,K)  in the 0.5-2 keV spectral range,  following prescriptions discussed in Paper I.

Each simulation of our grid of galaxy mergers and encounters starts
from systems built up with the 
initial conditions listed in Table~\ref{tab:tab1} and the parameters tuned as in the above cited papers, as described in \citet{Mazzei14a, Mazzei14b, Mazzei18}.  This set accounts for different  masses (from 10$^{13}$\,M$_{\odot}$ to 10$^{10}$\,M$_{\odot}$ for each system), mass ratios (from $1:1$ to $10:1$),  gas fraction (from 0.1 to 0.01), and particle resolution (initial number of gas and DM particles from 60000 to 220000).
By seeking to exploit a wide range of orbital parameters, we varied  the orbital initial
conditions in order to have the first peri-center separation, p,  equal to the initial length of the major axis of the more massive triaxial halo down to 1/10 of the same
axis,  for the ideal Keplerian orbit of two points of given masses. For each value of p we changed the eccentricity in order to have hyperbolic orbits of different energy. The spins of the systems are generally parallel to each other and perpendicular to the orbital (XY) plane. Misaligned spins have also been analyzed in order to investigate the effects of the system initial rotation on the results.
Table \ref{tab2} reports the initial conditions of each simulation which best reproduces, at a given snapshot, the global current properties of our targets.
 We recall here that by major mergers we mean mergers with the initial mass ratio of halo progenitors (Table~\ref{tab2})  $\leq 4$, while in minor mergers the ratio is  $> 4$.
In particular, col.5 provides the total initial  number of particles (the number of gas particles is the same as DM ones). The  initial gas mass resolution, m$_{gas}$, is between 1$\times$10$^6$ and 1.33$\times$10$^7$\,M$_{\odot}$. The star formation efficiency is 0.4 and the stellar mass resolution in each simulation ranges from 0.4$\times$m$_{gas}$ to 0.04$\times$m$_{gas}$ \citep[e.g.][]{Mazzei14a}, that is, for simulations in Table \ref{tab2},  goes from 0.40$\times$10$^5$  to 52$\times$10$^5$\,M$_{\odot}$. As a comparison, the mass resolution range of  \citet{EM18}  is 3.5-20$\times$10$^5$\,M$_{\odot}$.  Recent cosmological simulations of \citet{Kaviraj17}  and \citet{Martin18} have mass resolution of $\simeq$2$\times$10$^6$\,M$_{\odot}$.
The time step between individual snapshots is 37\,Myr. 
The gravitational softening is 1, 0.5, and 0.05\,kpc for DM,  gas, and star particles respectively.   
Our stellar spatial resolution, limited by  softening length, is  50\,pc, to be compared with the  200\,pc as  the best spatial resolution of \citet{EM18}  and $\approx$1\,kpc in \citet{Martin18}.
The final number of particles is at least twice that of the initial number,   ranging from  about 1.33 to 3.65$\times$10$^{5}$.
We point out that the resulting SFR,  the  driver of the evolution,  converges when the initial particle number is above 10$^4$   (MC03, their Fig.1, and \citet{Chri2010,Chri2012}).
The  main results are presented in the next section.  The details of the match of each target from 
the simulation which best reproduces its morphological, photometrical and kinematical properties are reported in Appendix.

\section{ Results.  }  
\label{Results}
\begin{figure*}[htb]
\begin{center}
\includegraphics[width=\textwidth]{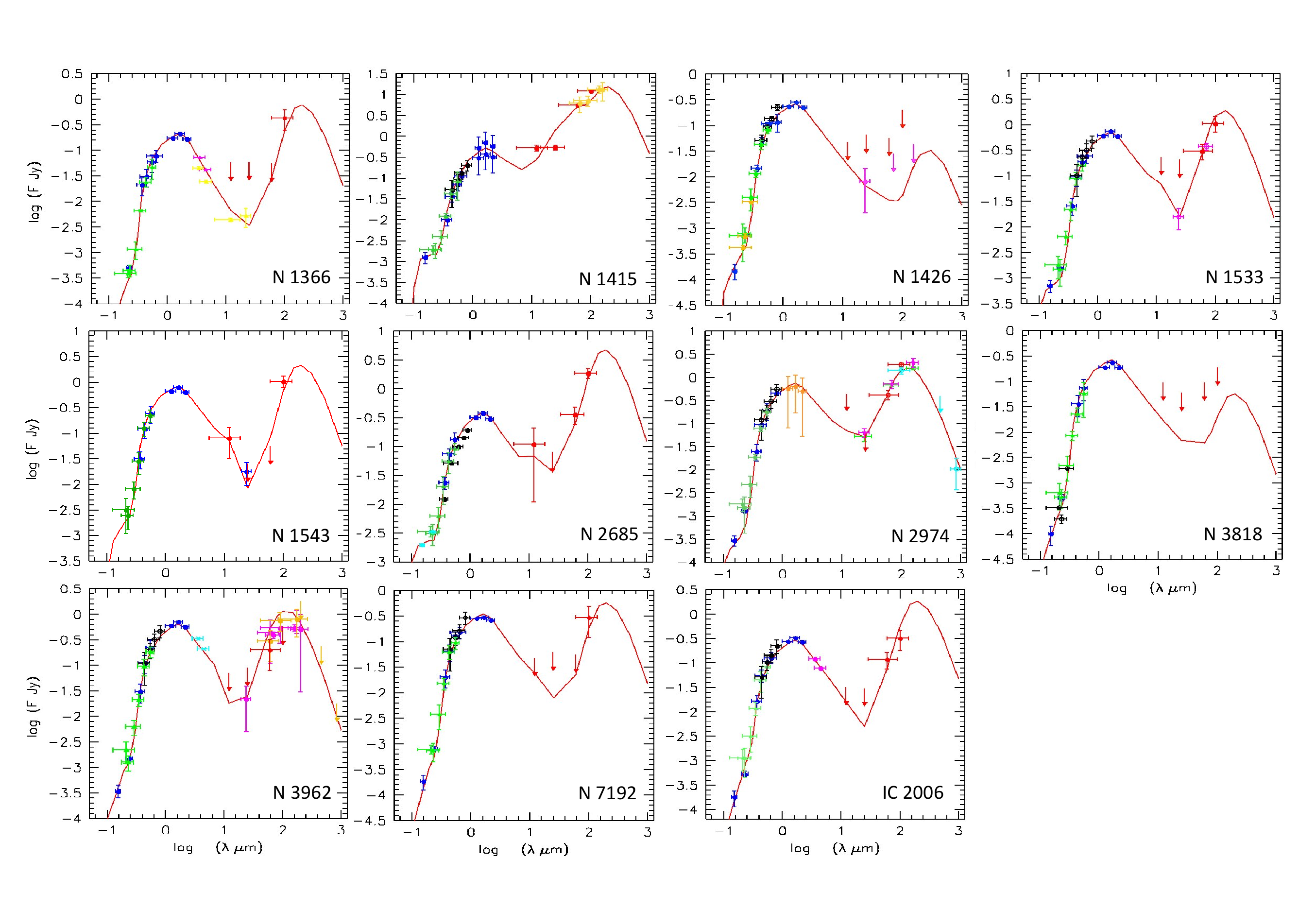}
\caption{The  SEDs of our targets (colored dots) compared with predictions from SPH-CPI simulations in Table \ref{tab2} (red solid lines) at the galaxy ages in Table \ref{tab3p}; green triangles are {\tt Swift-UVOT}  magnitudes from Paper II,  blue filled circles  are  from {\tt NED},  black  dots from CGS catalogue \citep{Ho11,Li11}, and red  points are {\tt IRAS}  fluxes from {\tt NED}; other details are in Appendix. Data are corrected for our own galaxy extinction following prescription of {\tt NED} and Paper II in the {\tt Swift-UVOT} filters.}
\label{SEDs}
\end{center}
\end{figure*}
\begin{figure*}[htb]
\begin{center}
\includegraphics[width=8.9cm]{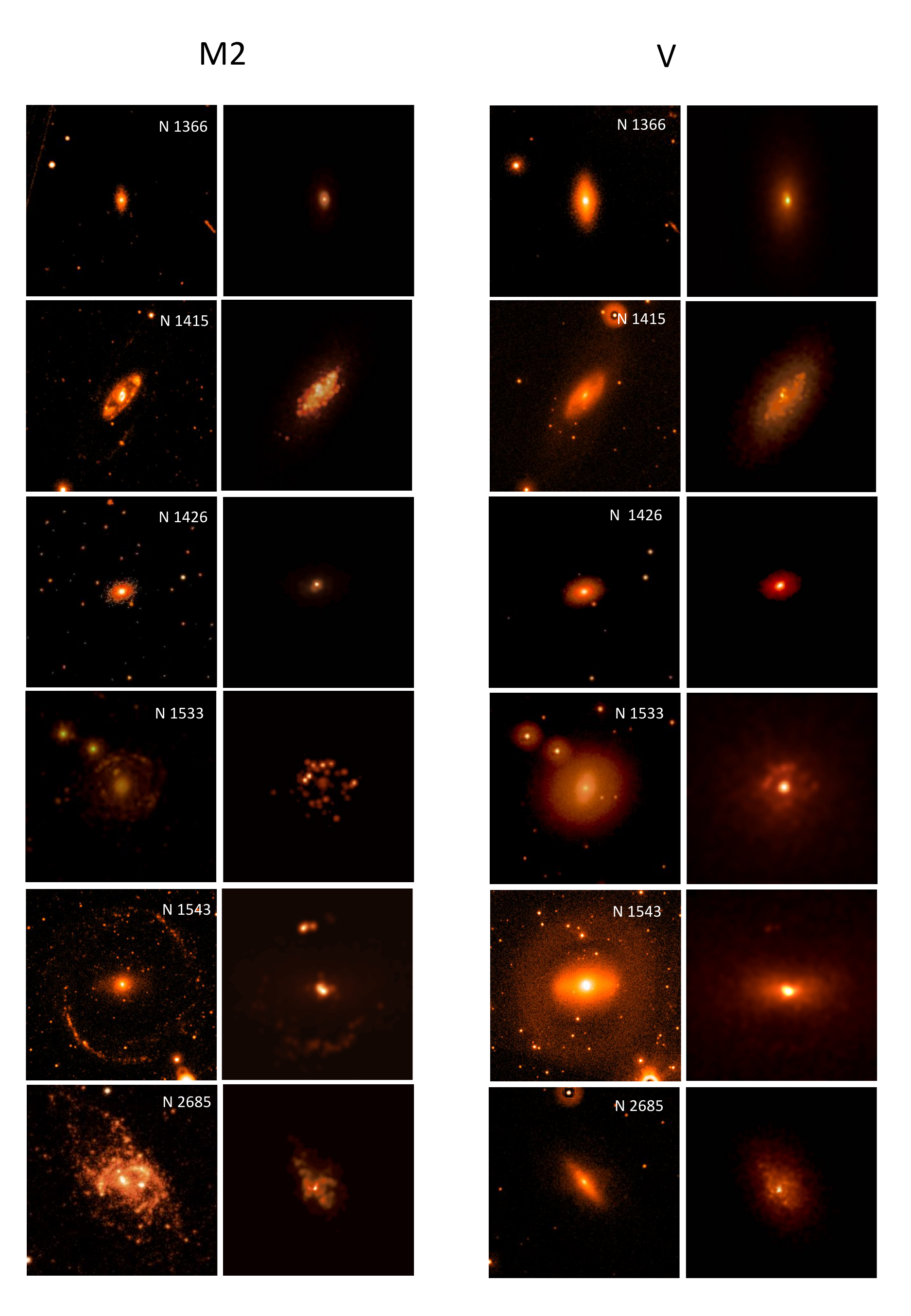}
\includegraphics[width=8.9cm]{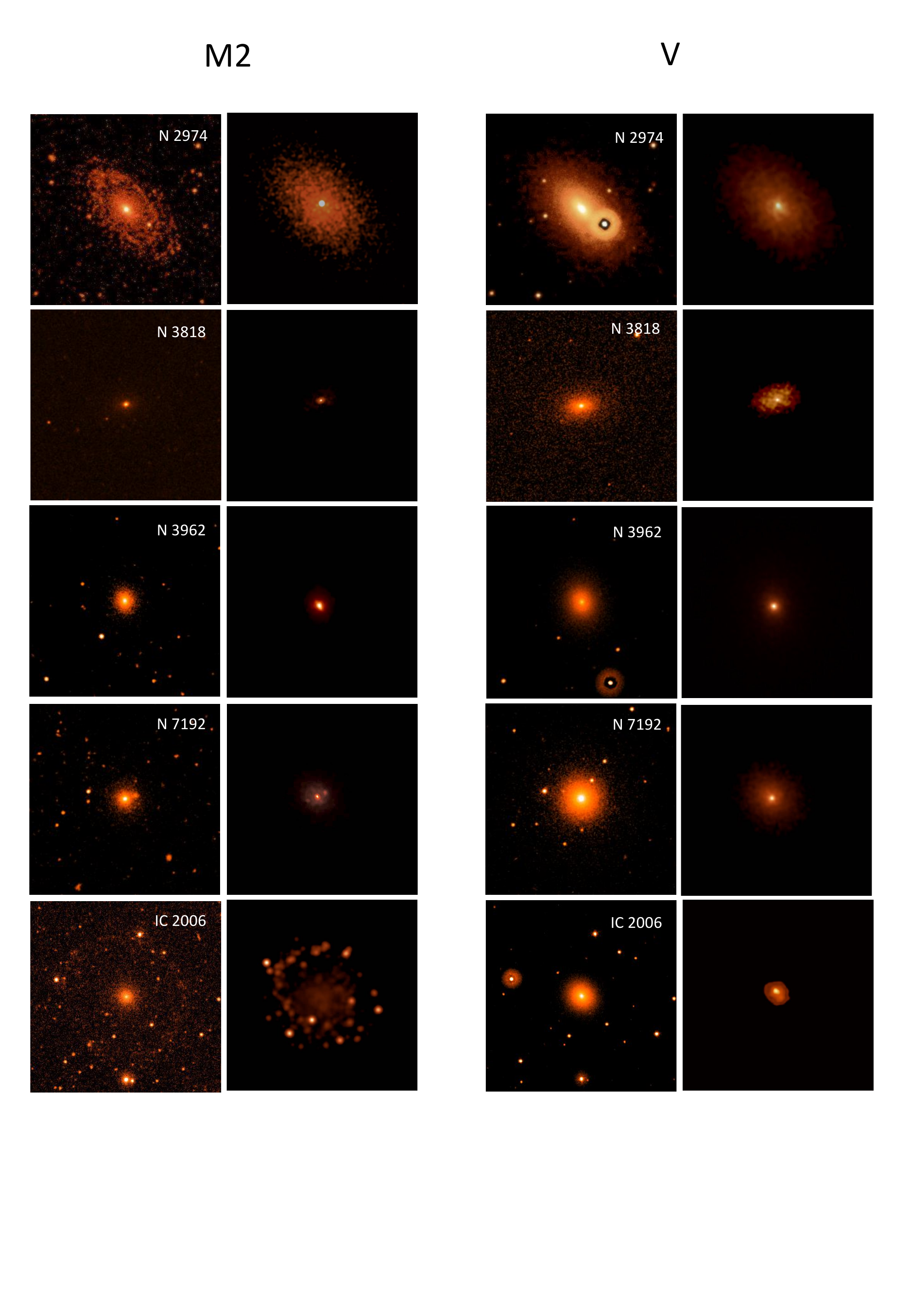}
\caption{Observed vs. simulated image of the 11 galaxies. In each column: {\it left}: the {\tt Swift-UVOT}  M2 (col.s 1, 5) or V (col.s 3, 7) band image; {\it right}: the simulated image (col.s 2, 6, and 4, 8), with the  same field of view and resolution (1\arcsec/px) of the observed one. All images are
5\arcmin$\times5$\arcmin, except for: NGC~1415 with 8\farcm6$\times$ 8\farcm6,  and IC~2006 and NGC~2685, with 7\arcmin$\times$ 7\arcmin.
}
\label{UV_OT}
\end{center}
\end{figure*}
\begin{figure*}[htb]
\begin{center}
\includegraphics[width=18cm]{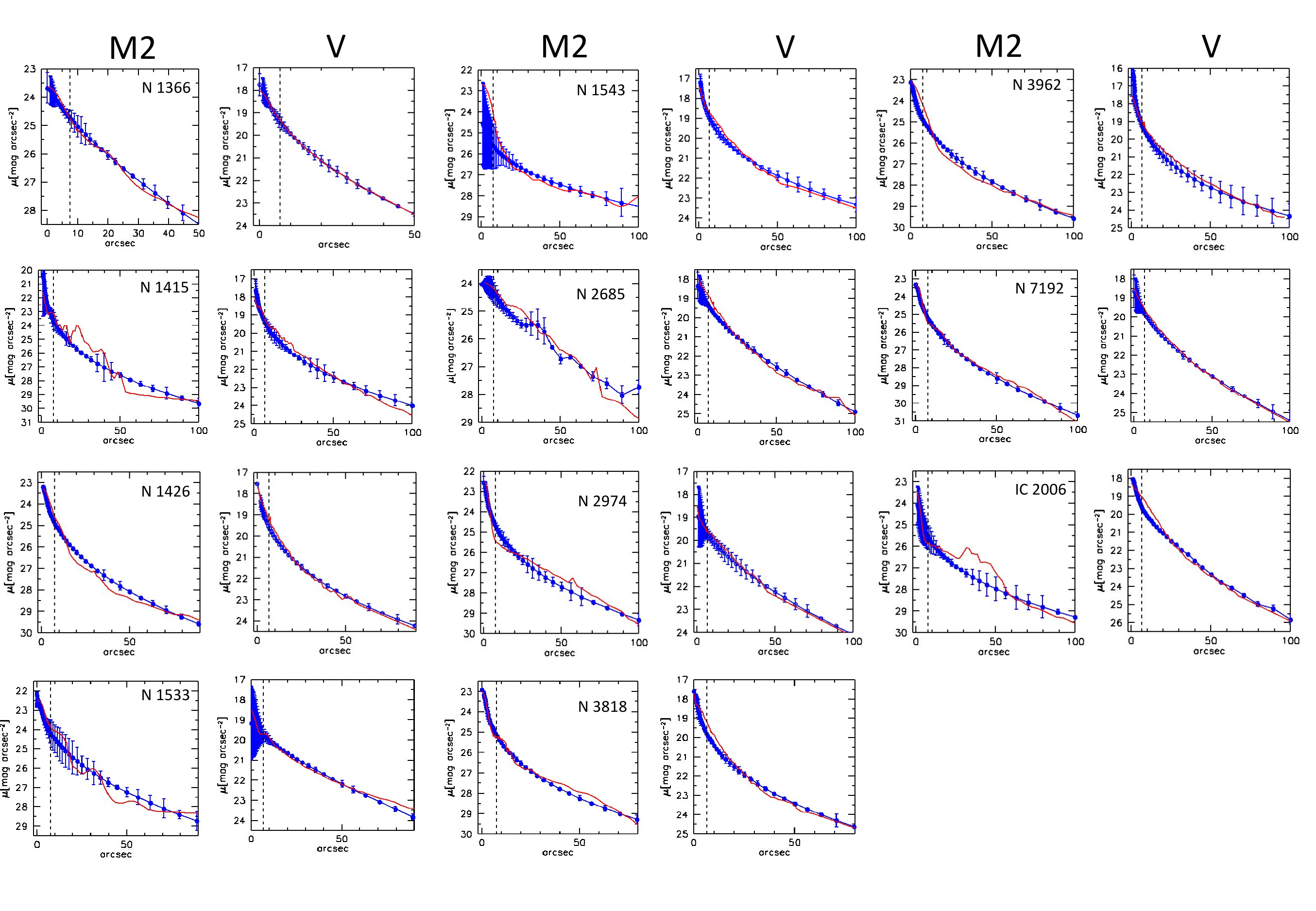}
\caption{Luminosity profiles of our targets in the  {\tt Swift-UVOT}  M2 and V bands;  blue dots are S\'ersic functions in Paper II, red solid lines  are from the simulation snapshots in  Fig.\ref{UV_OT}.}
\label{prof}
\end{center}
\end{figure*}

From our grid of SPH-CPI simulations we concentrate on the one  simulation that provides a snapshot which convincingly reproduces the 
global properties of the  ETGs considered, accounting  for the  following observational constraints: 
i)  total absolute B-band magnitude within the range allowed by observations (Table \ref{tab:tab1});
ii)  integrated SED as the observed one (cf. Fig. \ref{SEDs}); 
iii)  UV and optical morphologies (cf. Fig.\ref{UV_OT})  confirmed by a iv) comparison between  luminosity profiles (cf. Fig.\ref{prof}); v) kinematical properties in agreement with those in the literature; 
vi) X-ray luminosity of the hot gas,  and vii)   amount of cold gas consistent with the data in Table \ref{tab:tab1} 
 (see Appendix).
The results we present in Fig. \ref{ev_nof} and \ref{ev_f}, as we clarify below, are  evolutionary behaviors from the  simulation which best accounts for all the points above (i:vii) at the same snapshot.  This snapshot sets  the age of the galaxy accounted for from the onset of the SF. 
The galaxies studied span a large range of ages, between 9.6\,Gyr of NGC~3962, to 13.8\,Gyr of IC~2066, the oldest galaxy in our sample, (Table \ref{tab3p}) and  of total stellar masses, from 3.7 to 24.4$\times$10$^{10}$\,M$_{\odot}$ (Table \ref{tab3p}) with the most massive,  NGC~3962, being also the youngest.
The absolute magnitudes derived from the snapshots best fitting the global properties of our targets are  reported in Table \ref{tab3p}, and are to be compared with the observed ones in  Table \ref{tab:tab1}.  
Figure \ref{SEDs} allows the comparison between  the observed SEDs, extended over almost 4 orders of magnitude in wavelength, and  those derived from the selected snapshot (red solid line).   Error bars account for band width (x-axis) and 3$\sigma$ uncertainties of the flux (y-axis). 
The snapshot FIR SED is always composed by a warm and  a cold dust component both including PAH molecules as described in \citet{Mazzei92}, \cite{Mazzei94a}, and \citet{Mazzei94b}.
The warm dust component  is  heated by massive stars in HII regions,  the cold one by the diffuse light in the galaxy. 
UV and optical morphologies are compared in Fig.\ref{UV_OT}  with simulations in the same field of view and with the same resolution of the data. In order to derive the general  trend of the underlying  structure, the corresponding luminosity profiles are compared in Fig.\ref{prof}.
\begin{figure*}[htb]
\includegraphics[width=\textwidth, height=16cm]{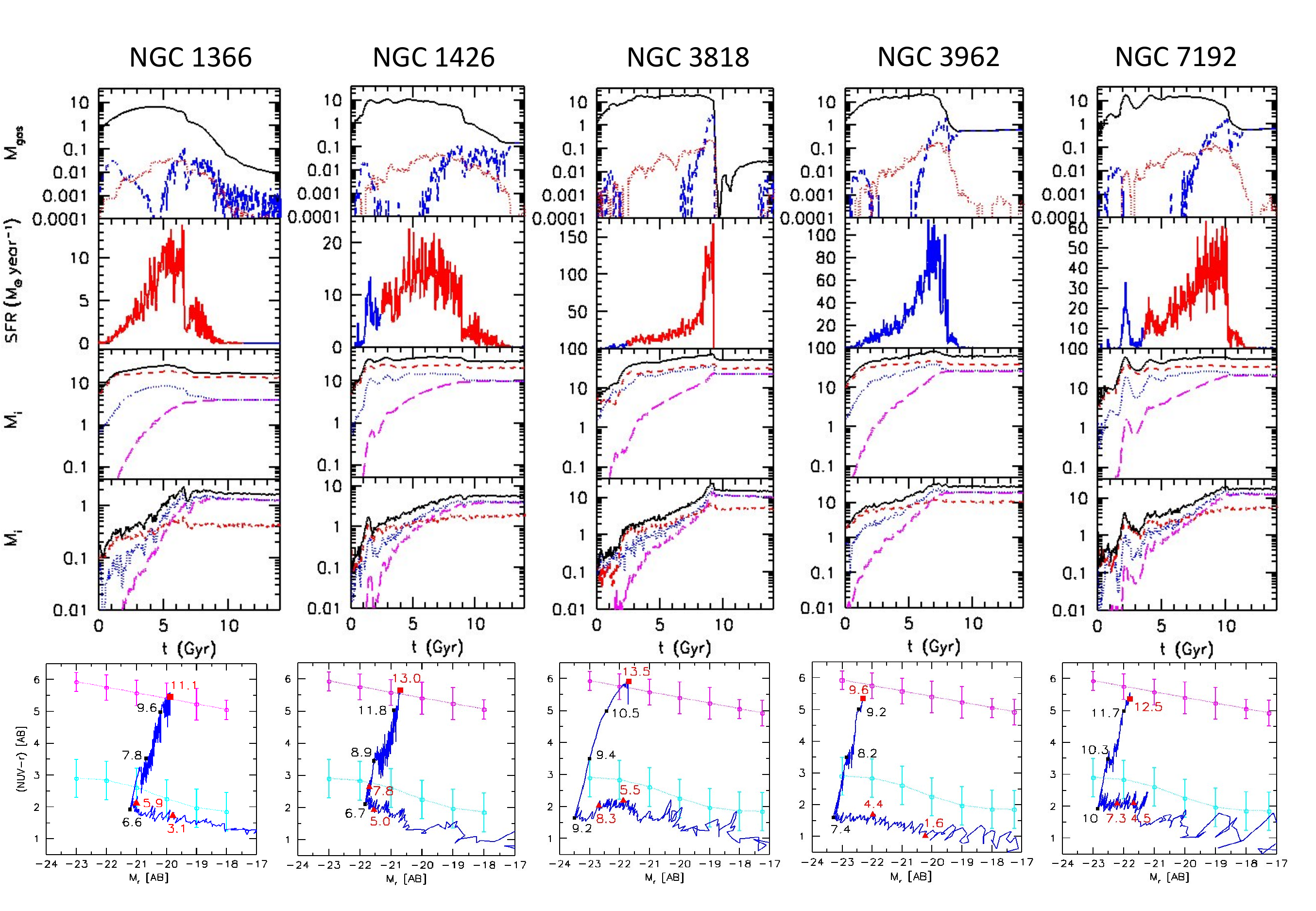}
\caption{ Evolutionary properties of ETGs without UV features.  From top to bottom: the gas accretion history, the SFR (red line from the beginning of the merger as defined in Section \ref{Results}),  the mass assembly history (see also Table \ref{tab5}) within 50\,kpc and R$_{25}$ in Table \ref{tab:tab1}, both centered on the B-band luminous center of the galaxy, and the CMD in the galaxy rest-frame;  more details in Section \ref{Results}.
M$_{gas}$ and the different mass components, M$_i$, are in units of 10$^{10}$\,M$_{\odot}$.}
\label{ev_nof}
\end{figure*}
Properties at points $v)$ to $vii)$ above are discussed  for each target in the dedicated section in Appendix.
Table \ref{tab3p} summarizes  the global properties of the simulations in Table \ref{tab2} from the snapshot best describing the  current  properties of our targets.

Relevant evolutionary properties are in Fig.\ref{ev_nof} and Fig.\ref{ev_f}. These include five panels for each target highlighting connections between general evolution and the path in the  CMD.  
This path   shows the behavior of the SFR,   how  each galaxy transforms and quenches. The NUV$-$r color is an excellent tracer of even small amounts of SF  \citep[e.g.][]{Mazzei14a}. 
The first three panels of  Fig.\ref{ev_nof} and Fig.\ref{ev_f}, from  top to  bottom, show  the gas accretion history,  the SFR and the mass assembly history derived within a fixed  reference radius of  50\,kpc, centered on the B-band luminous center of the galaxy.   The (black) solid  line in the top panels corresponds to the total mass of gas, the  (blue) dashed line  to gas with temperature   $\le$10$^4$\,K, and the (red) dotted line to gas with temperature $\ge$10$^{6}$\,K. 
Here, as in the following, we refer to cold gas as that with a temperature $\le$10$^4$\,K , given that its cooling time--scale is shorter than the snapshot time resolution (0.037 Gyr). This represents the upper limit of HI provided by each simulation.
The SFR is shown with a  blue solid line which  becomes red as  merger begins (when there is a merger), namely 
when the two stellar systems can no longer be distinguished because their  mass centers cannot be identified/separated anymore.
We find that the SFR  shows  a gentle self-quenching after reaching its maximum value, due to gas exhaustion and stellar feedback, lasting several Gyr in all our targets, with the exception of NGC~3818.  This galaxy reaches the highest value of the SFR among the targets examined (168\,M$_{\odot}$\,yr$^{-1}$, see Appendix).
The mass assembly history, that is the evolution of different mass components,  is shown within two fixed reference radii,  50\,kpc,  and  R$_{25}$  in Table \ref{tab:tab1} (panels three and four in Fig.s \ref{ev_nof} and \ref{ev_f}), to highlight their different behavior.
The (black) solid line is the total mass, the (red) short-dashed one  the DM,  the (blue) dotted line the gas+stars (baryons), and the (magenta) long-dashed one  the stellar component. 
The evolution is stopped at 14\,Gyr (c.f.g. Section \ref{Intro}) in all these panels.
We find that the DM mass exceeds always the stellar mass within the larger radius, and their mass ratio sets on a constant value while the SF quenches. 
An  opposite behavior occurs within the smaller radius, where the stellar mass exceeds the DM while the SF quenches.
The bottom panel presents the  path  of the galaxy in its rest-frame  optical--UV CMD.  In the following, as in \citet{Mazzei14a}, we assume that 
 the GV lies  between   NUV-r $= 3.5$, which marks the GV entry, and NUV-r $= 5$, the RS threshold. 
Significant evolutionary stages are outlined by large (black) dots; (red) triangles correspond to z=1 and z=0.5 using the  cosmological parameters in Section \ref{Intro}.  The red and blue local (z=0) sequences are reported following prescriptions in \citet{Wyder07}.  
Each galaxy moves along the BC until reaching its maximum  SF which corresponds to the brightest point  (BP in the following) of its CMD. Then, the SFR fades, and the quenching begins. This causes the crossing of  the BC and GV. The five galaxies of our sample without UV features all belong to the RS. Of the six showing  UV features, two: NGC~1415 and NGC~2685, belong to the GV. The evolution of NGC 1543, which belongs to the RS, shows short rejuvenation episodes in the last 2.3\,Gyr.   Rapid oscillations appear also along the path of NGC 1415 (see the SFR panels). Both NGC 1543 and NGC 1415 are minor mergers (Table~\ref{tab2}).
The  interaction/merger, occurred in the past of the galaxy,  produces the potential well and the gas reservoir which drive the gas assembly history. The growth of the stellar mass develops from the SFR driven by this accretion history.  The quenching, which gives rise to galaxy transformation, occurs from the behavior of the  SFR, that is gas exhaustion and  stellar feedback, several Gyr after the start of the interaction/merger.
Therefore, the merger or the interaction are not the main actual cause of the SF quenching. Rather, this is a consequence of the galaxy evolution and it is related to gas exhaustion and stellar feedback.
SNs feedback is enough to allow quenching of our targets whose  total stellar masses range  from 3.7 to 
24.4$\times$10$^{10}$\,M$_{\odot}$, while the total masses range from 16.8 to 61.2$\times$10$^{10}$\,M$_{\odot}$ (Table \ref{tab3p}).

\begin{deluxetable*}{ccccccc}[b!]
\tablecaption{Properties of eleven ETGs  from  simulations. \label{tab3p}}
\tablehead{
\colhead{Galaxy} & \colhead{ M$_B$}\phantom{000} & \colhead{age}\phantom{000} & \colhead{M$_*$}\phantom{000} &   \colhead{M$_{tot}$}  \\
  \colhead{}    & \colhead{(mag)}\phantom{000} &  \colhead{(Gyr)}\phantom{000}& \colhead{(10$^{10}$\,M$_{\odot}$)}\phantom{000}  &   \colhead{(10$^{10}$\,M$_{\odot}$)}\\
   \colhead{}    & \colhead{}\phantom{000} &  \colhead{}\phantom{000}& \colhead{M$_{R_{25}}$\phantom{0} M$_{50}$}\phantom{000}  &   \colhead{M$_{R_{25}}$\phantom{0}  {M$_{50}$}}
   \\
}
\startdata
NGC~1366  & -18.65 & 11.1  &  1.49\phantom{00} 3.68&2.06\phantom{00} 16.83  \\  
NGC~1415  & -19.30& 13.3 & 2.52\phantom{00} 5.39&4.99\phantom{00}  36.10 \\
NGC~1426 &-19.42&  13.0 & 3.75\phantom{00} 9.22 &5.79\phantom{00} 28.73 \\  
NGC~1533 &  -19.90&  13.7&7.16\phantom{0} 15.85&\phantom{0}10.64\phantom{00} 36.65\phantom{00} \\  
NGC~1543  &  -19.70&   10.7&  3.35\phantom{00} 8.60& 4.56\phantom{00} 25.70 \\  
NGC~2685 &  -19.72&  11.6&  4.05\phantom{00} 7.47&  6.21\phantom{00}  26.34\\ 
NGC~2974 & -19.92& 12.6& 6.90\phantom{0} 16.71& 10.47\phantom{00} 47.45\phantom{0} \\
NGC~3818 & -20.42& 13.5&  9.88\phantom{0} 21.38&15.01\phantom{00} 51.99\phantom{0}   \\  
NGC~3962 &  -21.24 & 9.6 &8.07\phantom{0} 24.43& 28.21\phantom{00}  61.23\phantom{0}\\  
NGC~7192 &  -20.52& 12.5 &11.80\phantom{0} 19.95\phantom{0}&17.82\phantom{0..} 52.84\phantom{0} \\ 
IC~2066    & -19.75& 13.8 &5.90\phantom{0} 18.07&  7.68\phantom{00}  39.26\\       
 \enddata
\tablecomments{
The intrinsic total B-band absolute magnitude is in col.1,  the galaxy age  in col.2, the stellar mass derived within our selected reference radii,  R$_{25}$ (Table \ref{tab:tab1}) and 50\,kpc in col.3, ({\it left} and {\it right} respectively), and the corresponding total mass in col.4, ({\it left} and {\it right}.) 
}
\end{deluxetable*}
\begin{figure*}[htb]
\includegraphics[width=\textwidth, height=16cm]{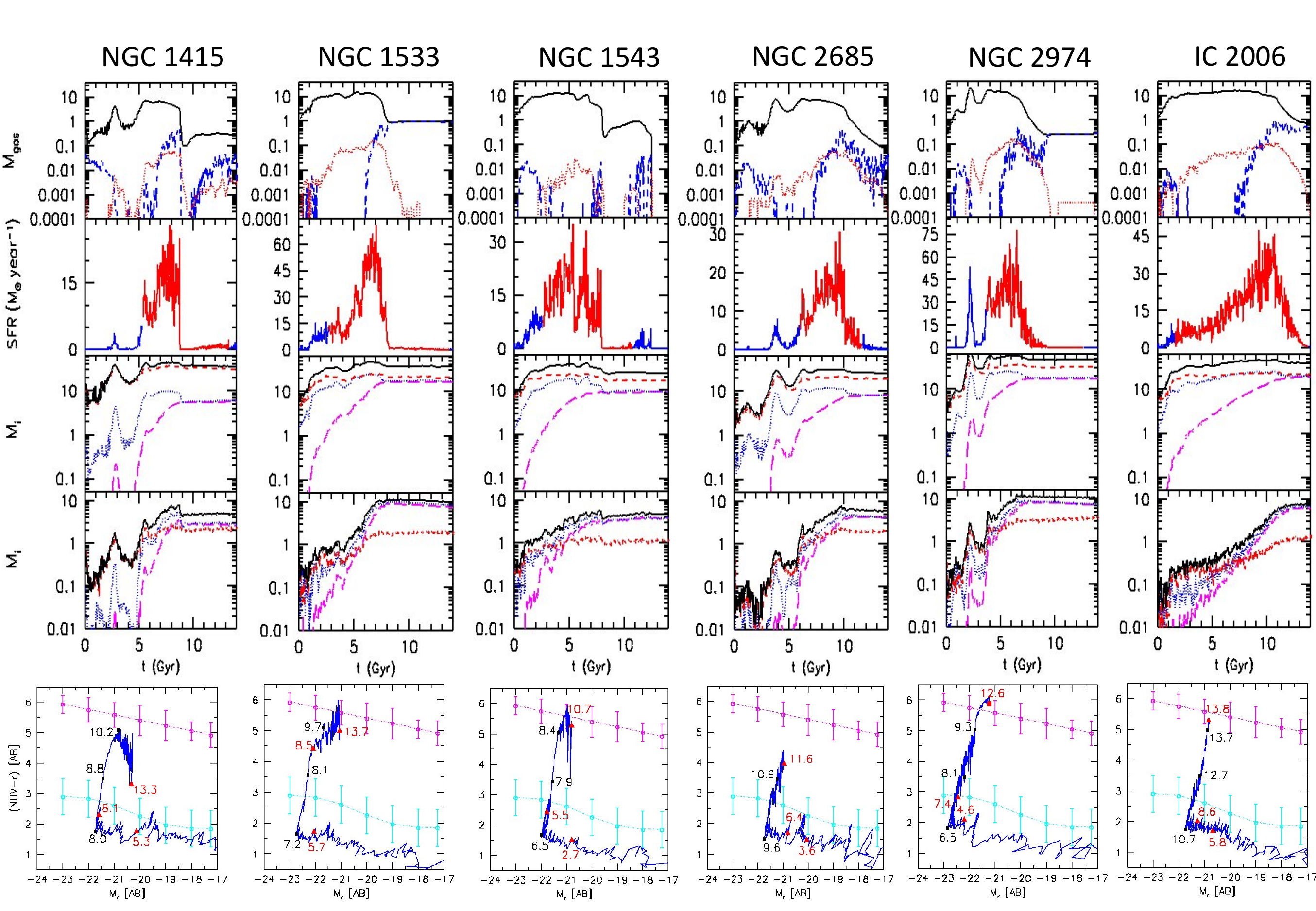}
\caption{As in Fig.\ref{ev_nof} for ETGs with UV features.}
\label{ev_f}
\end{figure*}
\begin{deluxetable*}{lrrrrrr}[b!]
\tablecaption{Percentages of the stellar mass growth in different evolutionary time/redshift intervals.\label{tab5}}
\tablehead{
\colhead{Galaxy} \phantom{000000}& \colhead{$\Delta$z(1--0)} \phantom{0000}& \colhead{$\Delta$z(0.5--0)}\phantom{0000} &  \colhead{$\Delta$z(BP--0)} \phantom{0000} &  \colhead{$\Delta$z(GV)}\phantom{0000}  & \colhead{$\Delta$z(GV--0)}\phantom{0000} \\ 
           \colhead{}            & \colhead{\%} \phantom{00}& \colhead{\%} & \colhead{\%}  &  \colhead{\%}  & \colhead{\%}
           }
\startdata
NGC~1366 \phantom{000}& 87.1\phantom{00} 96.3  & 31.7\phantom{00} 52.4   &  18.9 \phantom{00}14.1 &3.1\phantom{0} 12.0& 3.4\phantom{00} 10.3 \\  
NGC~1415\phantom{0000}& 87.0\phantom{00} 94.9 & 18.8\phantom{00} $-$3.2&21.7\phantom{000}  0.7&$-$1.2\phantom{} $-$4.2& 1.6\phantom{} $-$12.0 \\  
NGC~1426\phantom{0000}&  59.0\phantom{00} 79.4& 16.7\phantom{00} 34.6 &31.5\phantom{00} 53.0   & 3.8\phantom{00} 9.4 &4.3\phantom{00} 11.0 \\  
NGC~1533\phantom{0000}&  60.3\phantom{00} 86.8 & 0.5\phantom{00} $-$17.8 & 11.3\phantom{00} 11.7& $-$0.9\phantom{0} $-$3.8 &\phantom{i}0.7 \phantom{}$-$18.8 \\  
NGC~1543\phantom{0000}&   85.4\phantom{00} 89.3& 27.9\phantom{00} 49.0 & 8.0  \phantom{000}17.4 &$-$6.4\phantom{0} $-$2.9 & $-$2.1\phantom{0} $-$2.9 \\  
NGC~2685\phantom{0000}&  92.5\phantom{00} 99.6& 71.9\phantom{00} 94.0 & 15.9\phantom{00} 18.1& 1.6 \phantom{0}$-$0.7& 1.6\phantom{0} $-$0.7\\ 
NGC~2974\phantom{0000}&   62.4\phantom{00} 81.5 & 2.73\phantom{00} 7.34  & 14.1\phantom{000} 1.6  & 0.9\phantom{0} $-$1.0&\phantom{0}0.5\phantom{0}$-$14.9&  \\ 
NGC~3818\phantom{0000}& 78.7\phantom{00} 93.4&39.6\phantom{00} 58.6 & $-$4.9\phantom{0} $-$36.5  & $-$1.9\phantom{0} $-$0.7 & $-$3.0\phantom{0} $-$8.5 \\  
NGC~3962\phantom{0000}&98.2\phantom{00} 99.2&  82.9\phantom{00} 91.2  & 12.3\phantom{00} 11.3& 1.1\phantom{00} 4.5& 1.3\phantom{000} 4.1\\  
NGC~7192\phantom{0000}& 82.7\phantom{00} 94.3&  57.4\phantom{00} 79.6&3.8\phantom{00}  \phantom{00}0.2 &1.5\phantom{00} 6.4 & 1.8\phantom{000} 4.1 \\ 
IC~2066\phantom{0000} & 80.5\phantom{00} 97.9&  52.4\phantom{00} 88.9 & 10.3\phantom{00} 37.3&  0.7\phantom{0} $-$2.3&\phantom{0}0.9\phantom{0} $-$0.11\\       
\enddata
\tablecomments{
In col.s 2-- 6: percentages within  50\,kpc are on the left,  those within   R$_{25}$ (Table \ref{tab:tab1}) on the right.  BP is the Brighter Point  and GV the  Green Valley of the CMD of each galaxy. NB: the negative fractions indicate mass loss in the interval considered. 
} 
\end{deluxetable*}
\begin{figure*}[htb]
\begin{center}
\includegraphics[width=10cm]{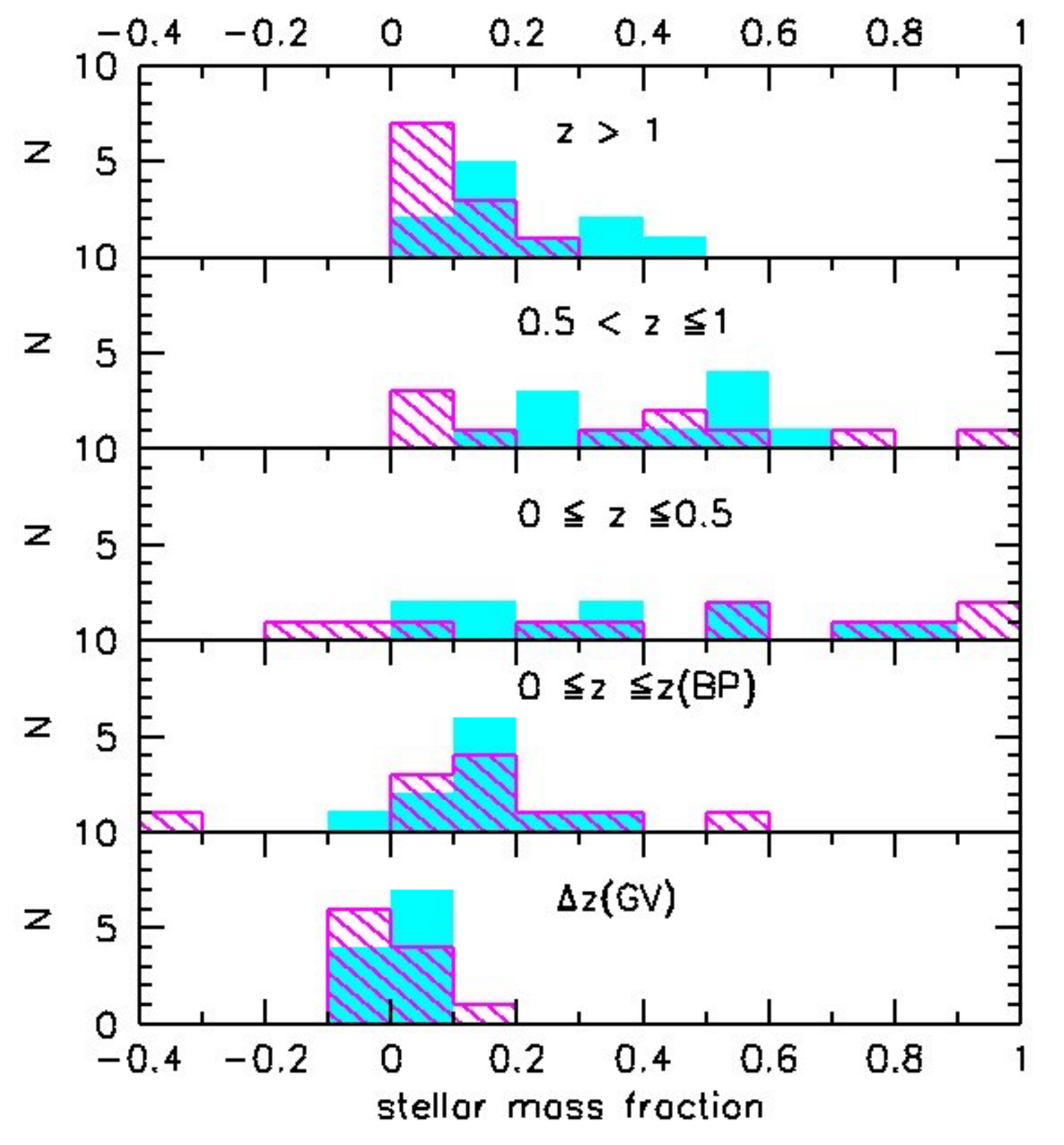}
\caption{Number of galaxies vs the fraction of  stellar mass assembled in different evolutionary steps within 50\,kpc, (cyan)  filled, and R$_{25}$ in Table \ref{tab:tab1}, (magenta tilted). Accounting for cosmological parameters in Section \ref{Intro},  the look-back times to redshift 0.5 and 1 are about 5.2 and 8\,Gyr respectively; BP and GV indicate  the Brighter Point  and the  Green Valley of the CMD of each galaxy. 
NB: the negative fractions indicate mass loss in the interval considered. 
}
\label{discuss1}
\end{center}
\end{figure*}
\section{Discussion}
\label{Discussion}
We find that there are no significant  differences in the formation mechanisms between ETGs with or without UV features.  Among those with UV features, NGC~1543 and NGC~1415 are minor mergers, and NGC~1533, NGC~2685, NGC~2974, and IC~2006 are major mergers.  Among the normal  (no UV features) ETGs, NGC~1366 and NGC~3962 are due to encounters, or  flyby,  with a very small companion (mass ratio 10:1),  and NGC~1426, NGC~3818, and NGC~7192 are major mergers.
Therefore,  major mergers reproduce well  the current global properties of ETGs both with long lasting (several Gyr) and without UV features, minor mergers  look more suited to ETGs with  short time ($\le$0.5\,Gyr) but long lived UV features, and  galaxy encounters   with a small companion match well normal ETGs.
Minor merger here analyzed   produce rejuvenation episodes which move  the galaxies from the RS back to the GV (Fig.\ref{ev_f}, bottom panels).   Moreover, while these galaxies cross the GV or return to the GV region, they are characterized by extended features in their UV morphologies.

Table \ref{tab5} and  Fig. \ref{discuss1} summarize  the growth of the stellar mass within two fixed reference radii (Section \ref{Results}) of   simulations in Table \ref{tab2}. We find  that 
 all  galaxies in our sample assembled  at least 60\% of their stellar mass  within 50\,kpc at z$\le$1.
At z$>$1, 45\% of our sample, namely  NGC~1366, NGC~1451, NGC~1543, NGC~2685, and NGC~3962, accreted less than 20\%  of their actual mass,  NGC~3818  and IC~2006, accreted about 20\%, and  NGC~1426, NGC~1533, and NGC~2974, about 40\%. 

The maximum SFR occurred more than  3\,Gyr ago in about 73\%  of our targets, and between 2\,Gyr and 3\,Gyr ago  in the remaining three cases, namely, NGC~2685 at z=0.15  (2\,Gyr ago),  NGC~3962 at z=0.17 (2.2\,Gyr), NGC~7192 at z=0.20 (2.5\,Gyr). 
In 4 galaxies this maximum  occurred  at z$>$ 0.5, that is more than 5.2\,Gyr ago.  These are NGC~1426 among UV featureless ETGs, and NGC~1415,  NGC~1533, NGC ~2974 among those with UV features. 

\begin{figure*}[htb]
\begin{center}
\includegraphics[width=17cm]{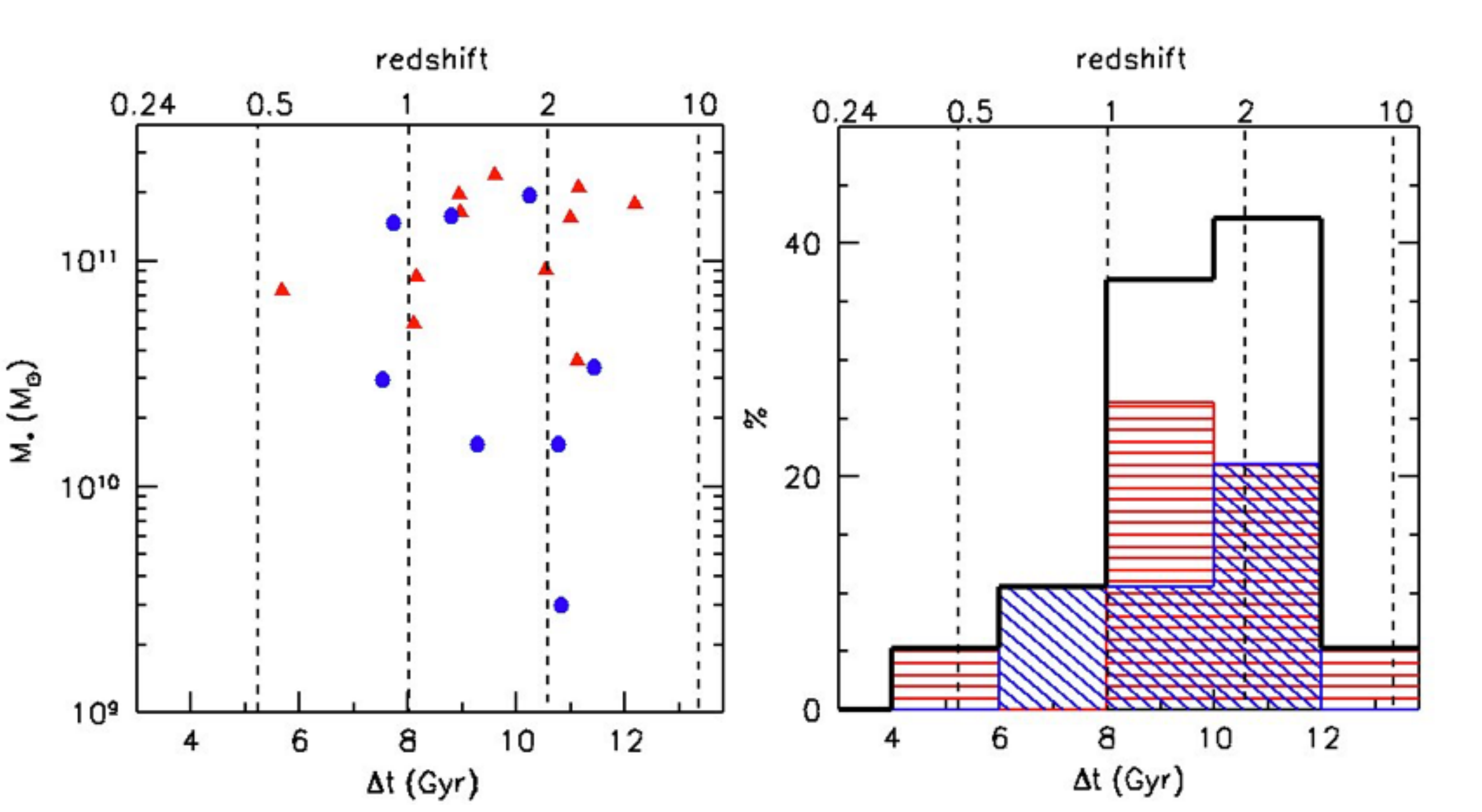}
\caption{ ({\it Left}) The total stellar mass at z=0  vs. age (either merger age or time of the onset of the SF due to galaxy flyby).  The merger age ($\Delta$t) is the difference between the  galaxy age at z=0 and that at the  beginning of the merger (Section \ref{Results}).
(Red)  triangles are  the eleven galaxies studied in this paper, (blue) circles are the eight galaxies from \citet{Mazzei14a}.  ({\it Right}) Histogram of the percentage of ETGs as a function of age (again, either merger age or time of the onset of the SF due to galaxy flyby). 
The horizontally shaded histogram is for galaxies in this paper, while  the  tilted  one is for those in \citet{Mazzei14a}, both  normalized to the total number of ETGs analyzed (19) (thick line). 
}
\label{discuss2}
\end{center}
\end{figure*}

\begin{figure*}[htb]
\begin{center}
\includegraphics[width=17cm]{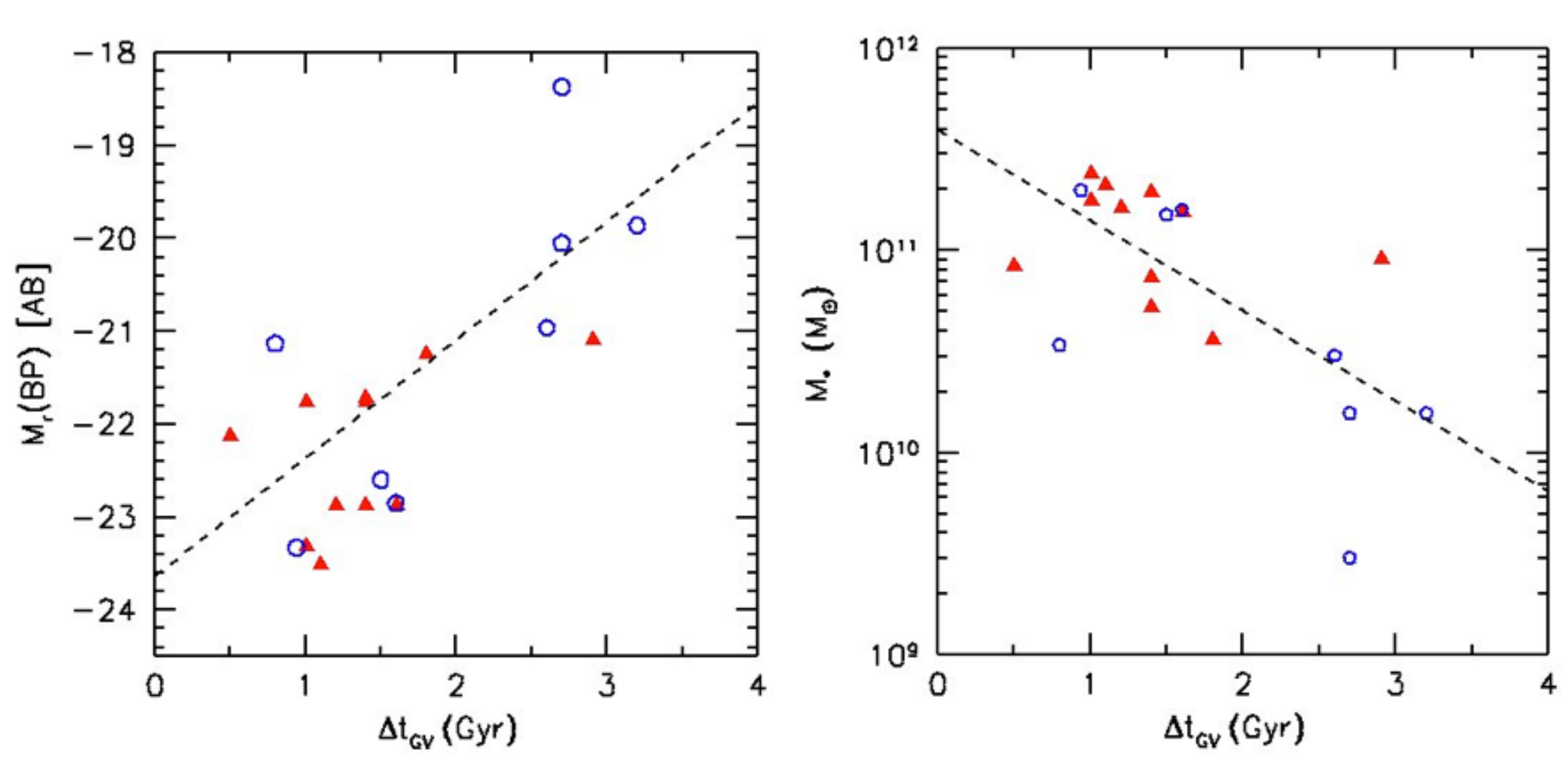}
\caption{ ({\it Left}) The absolute M$_r$ magnitude reached at the Brightest Point of the rest-frame CMD diagram  vs.
the  time spent in the GV, as defined in Section \ref{Results};  the correlation (dashed line) index is 0.72, and the slope of the regression is 1.27. 
({\it Right}) 
The  total stellar mass at z=0 vs.
the  time spent in the GV; the correlation (dashed line) index is -0.67, and the slope of the regression is -0.45.
Symbols are as in Fig.\ref{discuss2}.}

\label{discuss3}
\end{center}
\end{figure*}

At most 4\% of their total stellar mass was assembled after leaving the BC (GV-0 in Table \ref{tab5}).
As we noted above, NGC~2685 corresponds to a late merger,  occurred 6\,Gyr ago, at  z=0.61.
All the other mergers took place at  redshift $\ge$  1,  therefore  the age of the merger is  $\ge$8\,Gyr (Fig. \ref{discuss2}). 
In this figure, as in the following one (Fig.\ref{discuss3}), we also include 8 ETGs in two nearby groups, USGC~U367 and LGG~225 studied in  \citet{Mazzei14a}. We extracted simulations from the same grid for these further galaxies to match to their global properties.
Therefore, we increase the sample to 19 objects.  The span in stellar masses for the additional sample is 0.32--19.81$\times$10$^{10}$\,M$_{\odot}$, and in total masses is 2.78--63.47$\times$10$^{10}$\,M$_{\odot}$. Thus, the total range of stellar masses explored,  0.32--24.4$\times$10$^{10}$\,M$_{\odot}$, is about 2 orders of magnitude. 
Figure \ref{discuss3}  shows that the time spent by crossing the GV reduces with increasing  luminosity of the BP of the CMD diagram (left) and current stellar mass (right).  The correlation indexes are  0.72 and -0.67 respectively.
This indicates that  more massive galaxies supported higher SFRs in the past \citep[][and references therein]{MC03,Ro11},  and crossed the GV more quickly. Massive galaxies have undergone a faster morphological transformation from blue  to red  colors \citep[e.g.][]{Mazzei14b} than  less massive ones.  
Therefore, the evolution towards  ETG appears to be mass dependent.  The transit time of the GV traces this transition,  lasting some Gyr.
During this phase the SF is quenching and     the relaxed shape of these systems as we see them today is producing \citep{Mazzei14b}.
Recent results from dissipative merger simulations by  \citet{EM18},  which however start from
systems just formed,  composed of a spherical non-rotating  DM halo, and by a disk of gas particles with or without the presence of a stellar bulge, also predict that the length of the relaxation process is of some Gyr. 
The interaction/merger  occurred in the past of the galaxy, at redshift higher than 1 on average (Fig. \ref{discuss1}, right panel),  drives  the gas assembly history which regulates the SFR, and as a consequence the  galaxy evolution and transformation.  
Moreover, the  merger or the interaction are not the main reasons of the SF quenching.  Rather, this is a consequence of the  galaxy evolution and it is related to gas exhaustion  and stellar  feedback. This picture is consistent with recent findings by  \citet{Eales17a, Eales17b, Eales18},  \citet{Bremer18}, and \citet{Kelvin18}.  \citet{Eales17a} analyzed a volume limited sample of 323 nearby galaxies from the Herschel Reference Catalogue \citep{boselli10} 
    and investigate their SFR vs. mass properties using several indicators.
The sample is considered representative of the end-point of the galaxy evolution in the last 12\,Gyr in the local Universe. 
In their {\it Galaxy Main Sequence}   \citet{Eales17a}  plot  the galaxy mass vs. the specific 
SFR (SSFR, their Figure 2) showing that late-type galaxies (LTGs) and ETGs 
form a  sequence with negative slope without any distinct separation 
between LTGs and ETGs. They conclude that a rapid quenching process is not required. A more gentle
process, which they call  "slow quenching", better describes observations. In other
words, SF properties change gradually accompanied by a morphological
transformations. 
The evolution  that comes out of our SPH-CPI simulations  is consistent with this picture.
Quenching   is   faster in the more massive systems (see Figure~\ref{discuss3}).
Looking at  Figures \ref{discuss2}  and Table \ref{tab5},  we show that the merger  (defined in Section \ref{Results})  occurred at redshift $\ge$1 for the large majority of our sample (84\% of the larger sample considered, or 91\%, 10 out of 11 here), and that these galaxies assembled almost all their stellar mass  from the merger event to the present. Therefore, our galaxies evolve in rather different way from the 
assumption by \citet{Weigel17} which investigated the process of major merger quenching and its  importance  at z$\le$0.5. Our SPH-CPI simulations,  indeed, show that
while transitioning from the merger to the red early-type stage, galaxies  gain significant amounts of their stellar mass at odds with \citet{Weigel17} assumption. 
Their  analysis  suggests that major mergers are likely to lead to an evolution from star forming to quiescent  galaxies via quenching. However,  they specify that it is unlikely that major mergers  account for the majority of the quenching expected within the last 5\,Gyr. They further conjecture the 
existence of alternative quenching channels to explain the existence of the 
GV and RS population, which are likely to lead to a slow GV transition.
Our SPH-CPI simulations suggest that 
mergers and interactions  are important mechanisms to drive galaxy transformations, and  that
SF quenching due to gas exhaustion and SNs feedback, several  Gyr later the merger/encounter occurred, is enough for galaxies in low density environments. 
Morever,  quenching, which drives the crossing of  the GV, depends on the galaxy mass (Fig. \ref{discuss3}, left).
No alternative/additional quenching channels are required other than the gas exhaustion by SF and  stellar feedback. 

 Cosmological simulations have shown that AGN feedback is required to reproduce the cosmic SFR, down-sizing phenomena, and mass-redshift dependence of the specific SFR \citep[][and references therein]{Wang19}. At the same time, the cosmological simulations of \citet{TayKob15}  showed that galaxy evolution  in the stellar mass range  9--11 log(M$_{\sun}$) is almost unaffected by  AGN/BH feedback.Semi-analytic simulations by \citet{Henriques19}, that include AGN feedback, also found that SNs dominate galaxy quenching  in their intermediate mass halos regime.  Hydrodynamical simulations by \citet[][EAGLE project]{Katsianis17},  claim that SNs feedback is important at all redshifts and plays a major role in reproducing the observed number density of star forming galaxies at all epochs.
All these works point out the  role of SNs feedback in the galaxy quenching and thus in the galaxy transformation. We also emphasize that, in the range of total mass 10.4--11.8 log(M$_{\sun}$), 
and galaxy stellar mass,  9.5--11.4 log(M$_{\sun}$), 
 more massive galaxies are quenched  faster than less massive ones,  because gas is more rapidly transformed in stars, and SNs feedback is also enhanced.  As a consequence,  brighter,  massive galaxies spent a shorter time in the GV than the fainter, less massive ones  \citep{Mazzei14b}. The galaxy transformation to ETGs is faster in more massive galaxies. The quenching process is, however, a gentle process, lasting more than 1\,Gyr,
as  we found in NGC~3818, the most massive galaxy   by-product of a major merger we analyzed.
 Our findings are consistent with results of \citet{Genzel14}.  Their  percentage of  galaxies with  AGN at z$>$1 and log\,M $>$10.9\,M$_\odot$ is about 30\%.
The mass built up to z$>$1 of our simulated galaxies is, on average, less  than 20\% of their 
galaxy mass at z=0 (Table \ref{tab3p}). Only two systems have in place more than 40\% at z$>$1, 
(Table \ref{tab5}). Therefore, the galaxy stellar masses of our targets at z$>$1 are all below the threshold 
mass of \citet{Genzel14}.

Our conclusions are based on the results of  SPH-CPI simulations best describing the  current properties  of  19 ETGs in nearby groups.
The evolutionary picture we present agrees  with the recent findings by \citet{Eales17a} not only for discussion above concerning quenching length,  but also because several ETGs  analyzed here are not  ''dead" at  z$\le$0.5, as the authors themselves claim \citep{Eales17b,Eales18}.
Our simulations show  that the  percentage of  stellar mass  assembled within this redshift range  varies considerably from about 80\% to less than 1\%.
The stellar mass gained  while crossing the GV is, however,  always very small,   $\le$4\% (Table \ref{tab5}).

\section{Conclusions}
\label{Summary}

We use SPH-CPI simulations  anchored to the global properties  of 11 ETGs located in low density environments,   to shed light  on both the mechanisms at the origin of their UV  features, if any, and their evolution. All the simulations start from collapsing triaxial systems composed of DM and gas.  
We find that there are no significant  differences in the formation mechanisms between ETGs with or without UV features.
Major as well as minor mergers can trigger evolutionary phases  characterized by the appearance of  UV features sometimes extending up to the galaxy outskirts, as observed. In both cases, the UV morphology is long-lasting.  In minor mergers such as NGC 1415 and NGC 1543, each UV feature observed survives for less than 0.5 Gyr on average.  However, the UV emission is characerized by features that form   with continuity from the beginning of the merger to its current age, 
so that the overall morphology lasts 8.1 Gyr.  In major mergers, the 
resulting UV features last several Gyr; one of these features is the
inner ring of NGC 1533. 
Simulations of galaxy encounters  with a small companion match well the properties of normal ETGs.
Our set of galaxies spans a large range of ages, between 9.6\,Gyr  to 13.8\,Gyr  and of  total stellar masses, from 3.7 to 24.4$\times$10$^{10}$\,M$_{\odot}$  with the most massive,  NGC~3962, being also the youngest.

Our simulations, which include stellar feedback from SNs, show that the gas accretion history self-regulates the SFR which drives the galaxy evolution, without the need to invoke any external  or additional (e.g. AGN) quenching, as shown in Fig.s \ref{ev_nof} and  \ref{ev_f}, in agreement with the observing pictures of \citet{Bremer18} and \citet{ Eales17a, Eales17b,Eales18}. 
 Simulations by \citet{TayKob15},  EAGLE hydrodynamical simulations by \citet{Katsianis17} and recent results of semi-analytical simulations by \citet{Henriques19}, 
all outline the importance of SNs  feedback in driving the galaxy quenching, in suitable mass ranges.
Our results suggest that the AGN does not affect the general/global evolution (shown in Figs.~\ref{ev_nof} and~\ref{ev_f}) of our targets. 
However, AGN feedback could be important in the nuclear regions, at scales below those we have considered here (50 pc). 

The quenching is mass dependent.  Brighter, massive, galaxies spend a shorter time in the GV than the fainter, less massive ones. 
Both these conclusions are  based on the results of the SPH-CPI simulations presented here, anchored to the  current properties  of 11$+$8 ETGs in nearby groups, whose stellar mass range, 0.3--24.4 (10$^{10}$\,M$_{\odot}$),  is about 2 orders of magnitude. 
We plan to extend our work to a large number of ETGs, both to better sample  this stellar mass range and/or  to expand it.

The DM mass exceeds always the stellar mass within the larger radius we analyzed, 50\,kpc, and their ratio  tends to a constant value as the SF quenches. The opposite behavior occurs within the smaller radius, equal to the current value of R$_{25}$, for which the stellar mass exceeds the DM one as the SF quenches.
Mergers and/or interactions enhance the SF which proceeds differently from the case of a single, isolated, galaxy \citep{MC03} since  these mechanisms  deepen/modify the potential well where the gas is accreting. 
The growth of the stellar mass develops from the SF  driven by this accretion history. 
A large amount ($\ge$50\%) of the current stellar mass of each galaxy is still assembling at redshift $\le$0.5 for 35\% of our sample and would give a significant contribution to the IR bright galaxy population detected by {\tt Herschel}  \citep{Eales17a,Eales17b,Eales18},  {\tt ISO},  and {\tt IRAS} satellites \citep{Mazzei07}, before the complete morphological,  dynamic, and photometric transformation towards ETGs.


\acknowledgments

We acknowledge the usage of the {\tt HyperLeda} database 
(http://leda.univ-lyon1.fr) and  the NASA/IPAC Extragalactic Database (NED), which is operated by the Jet Propulsion Laboratory, California Institute of Technology, under contract with the National Aeronautics and Space Administration.
PM and RR acknowledge funding from the PRIN-INAF SKA 2017 program 1.05.01.88.04.
We thank the referee for helpful comments that helped improving our manuscript.
%






\appendix
\section{ETGs without UV features}
\label{Normal}
NGC~1366, NGC~1426, NGC~3818, NGC~3962, and NGC~7192  have no peculiar morphological features in their optical and UV images (Fig. \ref{UV_OT} and Paper II).

\subsection{NGC~1366}
\label{1366}
The simulation  which best describes the global properties of this galaxy corresponds to  an encounter between two counter-rotating halos with mass ratio 10:1,  our target being the more massive one  (Table \ref{tab2}).
The snapshot which matches the global properties of this S0 galaxy (Paper II) gives  M$_B$=-18.65\,mag (compare with Table \ref{tab:tab1}) at a galactic age of 11.1\,Gyr.
At this time the companion galaxy lies  656\,kpc  away from our target, about 107\arcmin  on the plane of the sky using cosmological parameters in Section \ref{Intro}.  Their closest approach occurred 5.9\,Gyr earlier, when our target was  5.2\,Gyr old, at z=0.61. 
The galaxy age  estimated by averaging those of  stellar populations at the best-fit snapshot is 7\,Gyr,  and  5.7\,Gyr  weighting these ages by B-band luminosity. These estimates agree well with  5.9$\pm$1\,Gyr  by \citet{Annibali07}. 
Figures \ref{SEDs}  shows the good match over almost four orders of magnitude  in wavelength between the snapshot SED  (red solid line) and the observations. 
Magenta points are from {\tt SG4}\footnote{
The Spitzer Survey of Stellar Structure in Galaxies (S4G), IRAC's 3.6 and 4.5 micron channels.
https://irsa.ipac.caltech.edu/data/SPITZER/S4G/}
 and yellow ones from the {\tt WISE} catalogue. 

The shape of the FIR SED is that expected  on average for ETGs \citep{Mazzei94a, Mazzei94b}, with a warm to cold dust ratio of 0.05 and  a cutoff radius of the cold dust distribution r$_d =$ 3\,kpc.  Moreover, the FIR to B-band luminosity ratio, 0.02, agrees  well  with the value
expected on average \citep{Mazzei94b}, with the FIR luminosity being 8\% of the bolometric luminosity.
Figures \ref{UV_OT}   and \ref{prof} show the good agreement of morphologies and luminosity profiles.
The X-ray luminosity from hot gas at this  age, 6$\times$10$^{37}$ erg\,s$^{-1}$, and the amount of cold gas (Fig. \ref{ev_nof}, top panel), %
agree with  upper limits in Table \ref{tab:tab1}. The central velocity dispersion, 115\,km\,s$^{-1}$,  fits well  the value of 113.9$\pm$3.8\,km\,s$^{-1}$ (HyperLeda catalogue), and the maximum stellar rotation velocity, 107\,km\,s$^{-1}$, agrees well with the value in Table 2 of Paper I (114$\pm$19\,km\,s$^{-1}$).
Looking at Fig. \ref{ev_nof},  the DM  exceeds the baryonic mass within the larger radius. Their mass ratio sets to a constant value, about 3.5,  starting from a galaxy age of 7.6\,Gyr. However, within the smaller radius, the stellar mass exceeds the DM by redshift 0.55, corresponding to a galaxy age of 5.5\,Gyr.  The stellar to DM mass ratio sets to a constant,  $\approx$3, starting from  8\,Gyr. 
At this age  the galaxy is leaving the BC to cross the GV  in its rest-frame  CMD (Figure \ref{ev_nof}, bottom). 
Table \ref{tab5} shows that this galaxy  assembles 87\%  and  32\% of its current stellar  mass from redshift 1 and 0.5, respectively,  but only 3.4\%
 in the past 3.3\,Gyr ,  that is from  z=0.27,  within   50\,kpc. In the same time/redshift steps,
the percentage of stellar mass growth is higher within the smaller radius, given the smaller stellar mass included (Table \ref{tab5}).

\subsubsection{NGC~1426}
\label{1426}
The simulation  which matches the global properties of this Elliptical is a major  merger (Table \ref{tab2}) which 
 begins  2.5\,Gyr after the onset of the SFR  (Section \ref{Results}).
The best snapshot corresponds to a galactic age of 13\,Gyr and M$_B$=-19.42\,mag.  
The age  of the galaxy is  9\,Gyr by averaging the ages of its stellar populations,  and   7\,Gyr  weighting these ages by B-band luminosity (Table \ref{tab3p}).  These estimates are in agreement  with the value of 9.0$\pm$2.5\,Gyr by \citet{Annibali07}.
Figure \ref{SEDs} compares the observed SED with that provided in Table~\ref{tab3p}.
In the UV spectral range, blue filled circles are  {\tt GALEX}  from \citet{Marino11a},  yellow points  from \citet{Hodges2014}, and, in the far-IR spectral range, magenta points  are from \citet{Temi2009}.
The FIR luminosity accounts for only 0.4\%  of the bolometric luminosity. The  shape of the FIR SED is  that expected on average for ETGs \citep{Mazzei94a, Mazzei94b}, with a warm to cold dust ratio of 0.5 and  a cutoff radius of the cold dust distribution, r$_d$,  of 3\,kpc.
The  X-ray luminosity from hot gas, 2$\times$10$^{37}$ erg\,s$^{-1}$,  agrees with the upper limit in Table \ref{tab:tab1}.
Both the central velocity dispersion, 132\,km\,s$^{-1}$,  and the maximum stellar rotation velocity, 97\,km\,s$^{-1}$,  are   consistent with the values in the HyperLeda catalogue,  146.7$\pm$1.9\,km\,s$^{-1}$ and, 114$\pm$19.0\,km\,s$^{-1}$, respectively.
Looking at  Fig. \ref{ev_nof},  the DM  exceeds the baryonic mass within  50\,kpc  (left) and their mass ratio sets a constant value, about 2.2,  from a galaxy age of 9.4\,Gyr. However, within R$_{25}$ (right) the stellar mass exceeds the DM one by redshift 0.75, corresponding to a look-back time of 6.8\,Gyr (i.e.  a galaxy age of 6.2\,Gyr).  The stellar to DM mass ratio stays constant, at  $\approx$2, starting from a galaxy age of 9.7\,Gyr. 
This simulation assembles  about 59\% and 17\%  of its stellar mass from z=1 and z=0.5, respectively considering 50\,kpc (Table \ref{tab5}), and no more than 4\%  by crossing the GV, accounting for several oscillations. The galaxy spends 6.3\,Gyr from the maximum value of its SFR, that is the BP point of its CMD diagram, and the following quenching (z=0) assembling  31.5\% of its stellar mass  within the same radius (Table \ref{tab5}).


\subsection{NGC~3818}
\label{3818}
The simulation  which best describes the global properties of this E galaxy is a major merger  (Table \ref{tab2}).
The best snapshot representing the global properties corresponds to a galaxy age of 13.5\,Gyr, with a B-band absolute magnitude of  -20.42\,mag    (Table \ref{tab3p}).
The  age of the galaxy becomes 8\,Gyr by averaging those of its stellar populations,  and reduces to 6.3\,Gyr  weighing these ages by B-band luminosity (Table 3).  These estimates agree with  \citet{Annibali07}  which give 8.8$\pm$1.2\,Gyr for this galaxy. 
Figures \ref{SEDs}--\ref{prof} compare observed and simulated  properties.
The blue filled circles in the UV spectral range are {\tt GALEX} data from \citet{Marino11a} and  (black)  dots  from \citet{Hodges2014}. 
The FIR SED,  not strongly constrained by the observations, corresponds to a FIR to bolometric luminosity ratio of 0.008, and to a FIR to blue luminosity ratio of 0.005,
in agreement with findings by  \citet{Mazzei94b} for a complete sample of Es.
The   X-ray luminosity of the hot gas expected at the best-fit  age agrees with that in Table \ref{tab:tab1},  being $\simeq$2$\times$10$^{39}$ erg\,s$^{-1}$.
Both the maximum velocity of the stars,   99\,km\,s$^{-1}$, and their central dispersion velocity, 180.3\,km$^{-1}$, are in good agreement with  observational constraints (Paper I, their Table 2) in the Hyperleda catalogue (106.7$\pm$14.8 and 193.4$\pm$3.7\,km\,s$^{-1}$, respectively).
Figure  \ref{ev_nof} shows that the SFR provided by this simulation reaches the highest value, 168\,M$_{\odot}$\,yr$^{-1}$, for all the simulations presented here.
Therefore, is it not surprising that  the residual SFR  cannot compensate the death rate during crossing of the GV (Table \ref{tab5}).  
The  galaxy moves along the BC of its rest-frame CMD until reaching  its highest SF and brightest magnitude, when the galaxy is  9.2\,Gyr old. Then, the SF quenches by itself, due to the gas ejection and exhaustion produced by its  high regime.  When the SF is stopped,  the stellar mass assembly becomes negative (Table \ref{tab5}). 
 The percentage of its current stellar mass assembled   from redshift 1 is  79\%, and  from z=0.5   40\%,  within 50\,kpc (Table \ref{tab5}).
 The DM  exceeds the stellar mass within this radius along all the evolution. Their mass ratio reaches a constant value, $\simeq$1.4, starting from about 10\,Gyr.
 Looking at the smaller fixed radius,  R$_{25}$  in Table \ref{tab:tab1} (Fig.\ref{ev_nof}), about 35\% of the current stellar mass is assembled  between redshift 1 and 0.5, and $\simeq$59\%  between 0.5 and 0.
The stellar mass exceeds the DM one starting from 8.4\,Gyr,  and their mass ratio becomes almost constant (2)  starting from 9.6\,Gyr, that is, in the GV.

\subsection{NGC~3962}
\label{3962}
This elliptical (E1) galaxy is a particularly intriguing case. Our UV  structural analysis (Paper II) does not  give any clear evidence of  a disk, however signatures of some recent SF have been detected  (e.g. Paper II). 
Moreover, this system entails a large   amount of  both cold  and hot  gas (Table \ref{tab:tab1}).
The simulation  which best reproduces the global properties of this galaxy corresponds to a flyby between two halos with mass ratio 10:1  (Table \ref{tab2}),  our target being the more massive one. From the snapshot  selected, the galaxy age is  9.6\,Gyr and the  B-band absolute magnitude, M$_B$=-21.2\,mag (Table \ref{tab3p}).  At this time the companion galaxy lies  513\,kpc away from our target,  about 50\arcmin\, in projection on the sky.  The closest approach, 507\,kpc, between the two galaxies is realized  0.37\,Gyr after the best snapshot.
The age of the galaxy, estimated by averaging the ages of its stellar populations, is 4\,Gyr  and reduces to 3.2\,Gyr  weighing their ages by B-band luminosity within R$_{25}$.
\citet{Annibali07}   estimate for this galaxy 10.0$\pm$1.2\,Gyr   whereas  \citet{Serra10} give 2.5\,${^{+0.4}_{-0.3}}$\,Gyr. 
This value is  in good agreement  with the central age (within 1\,kpc)  we derive by  weighing the ages of stellar population by B-band luminosity, that is 2\,Gyr. 
Figure \ref{SEDs} shows the match of the SED from the best snapshot  with the observed data.
Blue filled circles in the UV range are {\tt GALEX} data from \citet{Marino11a};   in the FIR spectral range magenta dots are ISO (Infrared Space Observatory) observations  from \citet{Temi2004, Temi2009},  and yellow triangles are  MIPS and SCUBA  upper limits at 450 and 800\,$\mu$m from  \citet{Leeuw2004}.
The FIR SED entails  about 9\% of the bolometric luminosity. 
The diffuse radiation due to cold dust  corresponds to a disk distribution of dust and stars with a central intensity double than that in our own Galaxy \citep{Mazzei92}. The warm dust component requires a temperature of 48.5\,K instead of 62\,K in the Milky Way, and a warm/cold luminosity ratio of 0.5. H$_{\alpha}$  observations by \citet{Z96}  highlighted the presence of a inner disk with a radius of about 1\,kpc (~5\arcsec), and a brighter arm-like feature extending up to  3.4\,kpc  (20$\arcsec$). Unfortunately, neither {\tt GALEX} \citep{Marino11b} nor {\tt Swift} (Paper II)  detected  UV features associated with this  emission.
This simulation provides a gas amount  in agreement with the  value in Table \ref{tab:tab1} (Fig.\ref{ev_nof}). 
This gas concentrates in the inner 4\,kpc,  in good agreement with  findings by   \citet{Z96}. Moreover, its maximum rotation velocity, 110\,km\,s$^{-1}$, agrees with the value provided by {\tt Hyperleda} database (99.1$\pm$7.7 km\,s$^{-1}$)  as well as the central (r$\le$1\,kpc) velocity dispersion of stars, 220.6\,km\,s$^{-1}$, comparable to 220.3$\pm$12.6 \,km\,s$^{-1}$. 
The X-ray luminosity of the hot gas 
expected at this  age is  $\simeq$1.5$\times$10$^{39}$ erg\,s$^{-1}$, about half that in Table \ref{tab:tab1}.
Figure \ref{ev_nof} shows that the DM  exceeds always  the stellar mass within 50\,kpc. Their mass ratio is  constant, 1.5, starting from  8.4\,Gyr. However, within the smaller radius, R$_{25}$, the stellar mass exceeds the DM one from 6.8\,Gyr, and their mass ratio becomes constant, 1.8, from  8.4\,Gyr again, when the galaxy is moving in the GV.
Following this simulation, the galaxy lives on the BC  7.4\,Gyr, until it reaches  its brightest magnitude. Then, the SFR quenches by itself due to the gas exhaustion produced by its  high regime and related feedback,  in about  0.8\,Gyr,  leading  the galaxy out of  the BC. The galaxy crosses the GV in 1\,Gyr  accounting for some  percent of its final stellar mass assembly within both our reference radii (Table \ref{tab5}). 
 
\subsection{NGC~7192}
The simulation which best-fits  of the global properties of NGC~7192 is a major merger  (Table \ref{tab2}). The galaxies merge  3.6\,Gyr after the onset of the SF.
The snapshot which matches  the global properties of this galaxy  provides a galaxy  age of 12.5\,Gyr and M$_B$=-20.5 mag (compare with Table \ref{tab3p}).
By  averaging  the age  of its stellar populations, we derive 7\,Gyr  and  $\simeq$5\,Gyr  weighing these ages by B-band luminosity, in good agreement with \citet[][and references therein]{Marino11a}.
In Figure \ref{SEDs}, which compares  the observed SED with that from the selected snapshot,
blue filled circles in the UV spectral range are {\tt GALEX} data from \citet{Marino11a}. 
Comparison of morphologies, and  luminosity profiles  are in Fig.\ref{UV_OT} and Fig.\ref{prof}.
The   FIR SED is that described  in  \citet{Mazzei94a},  with, however, a warm/cold dust ratio,  r$_{w/c}$=0.05. The FIR luminosity is equal to ~3.6\%  of the (intrinsic) bolometric luminosity and the FIR to blue luminosity ratio is 0.009,  in agreement with results of  \citet{Mazzei94b}.
The amount of cold gas in Table \ref{tab:tab1} agrees  with results from the selected snapshot, although a larger gas reservoir is expected from this simulation (Fig. \ref{ev_nof}, top).
Both the central velocity dispersion, 171\,km\,s$^{-1}$, and the rotation velocity  of the stars, $\le$20\,km\,s$^{-1}$,  agree with observational constrains, 177$\pm$5.4 and 0 \,km\,s$^{-1}$  respectively from the Hyperleda catalogue.
By exploiting {\tt Spitzer} IR spectra, \citet{Rampazzo13} found PAH signatures in the inner region of this system, pointing at an active stage of  SF in the past 2\,Gyr,  in agreement with the evolutionary picture highlighted in the CMD of Fig. \ref{ev_nof} (bottom panel). Moreover, they also emphasized a  possible non-thermal contribution due to LINER activity. 
Our simulation provides an  X-ray luminosity of the hot gas in this system  5 times lower than in Table \ref{tab:tab1}. 
The DM mass exceeds always the stellar mass within 50\,kpc. Their mass ratio becomes constant, 1.7, from 10.7\,Gyr, when the galaxy lies in the GV.  The percentages of stellar mass  assembled from redshift 1 and 0.5  are  about 83\%,  and  57\% respectively (Table \ref{tab5}).
Looking  within the smaller radius,  R$_{25}$ in Table \ref{tab:tab1}, the stellar mass dominates  the DM  starting from $\approx$8\,Gyr, corresponding to a look-back time of 4.6\,Gyr  (z=0.42, Section \ref{Intro}).  Their ratio sets on a constant value,  2, from an age of 10.4\,Gyr. 
 94\% of the current stellar mass is assembled from redshift 1 and $\approx$80\% from 0.5  within this radius (Table \ref{tab5}).
This galaxy lies  on the BC 10\,Gyr,  taking only 0.3\,Gyr to get out. Then, it crosses the GV in 1.4\,Gyr,  attaining its actual position in  2.2\,Gyr. The stellar mass assembled within 50\,kpc  during this route, that is from the GV to z=0,  is only 1.8\% of its current mass, and 4.1\% within R$_{25}$ (Table \ref{tab5}).\\


\section{ ETGs showing UV ring/arm like features}  
\label{features}
NGC~1415, NGC~1533, NGC~1543, NGC~2685, NGC~2974, and IC~2006,  show UV features (Fig. \ref{UV_OT}).
We point out that   S\'ersic functions, blue lines in Fig. \ref{prof},  cannot  account for  these features which appear, indeed,  in the UV luminosity profiles of simulations
(red lines).

\subsection{NGC~1415}
The global properties of this galaxy  are well matched by a minor merger  (Table \ref{tab2}).  Following this simulation NGC~1415 is 13.3\,Gyr old, with M$_B$=-19.3 mag (compare with Table \ref{tab:tab1}).
By  averaging  the age  of its stellar populations, we estimate 8\,Gyr  and   5.3\,Gyr  weighing the same ages by B-band luminosity.  Blue filled circle in the UV range  of Fig.\ref{SEDs}  is {\tt Galex}   FUV  from \citet{Buat07},  the higher fluxes in the NIR range are from \citet{spinoglio95}, and 
yellow data  in the FIR spectral range come from  AKARI FIS catalogue. 
The  FIR SED 
requires a warm to cold luminosity ratio,  r$_{w/c}$,  of 1.2,  more than twice as much as expected on average \citep{Mazzei94b}.
Moreover,  74\% of the intrinsic bolometric luminosity of this galaxy comes out in the FIR spectral range. These points contradict its classification as ETG, enforcing that of a late type galaxy (SB, see Table 1 in Paper II and Section \ref{Sample}).
The  amount of cold gas  and  the X-ray luminosity of the hot gas  in Table \ref{tab:tab1}  agree well with the results from the selected snapshot, 1.8$\times$10$^{9}$\,M$_{\odot}$ and  2$\times$10$^{39}$\,erg\,s$^{-1}$
respectively. Moreover,  the maximum rotation velocity of cold gas, 168\,km\,s$^{-1}$, is like the value in the Hyperleda catalogue (163.6$\pm$ 6.1\,km\,s$^{-1}$).	 

Figure \ref{ev_f} highlights  the evolutionary properties of the selected simulation.
The merger begins  (Section \ref{Results}) after 5.2\,Gyr (z=1.03) from the onset of the SFR.
Looking at the evolution of the different mass components within the  reference radii in Section \ref{Results}, 
a large percentage of the current stellar mass is assembled from redshift 1 within both these radii (Table \ref{tab5}).
The stellar mass  becomes equal to  the DM one  within R$_{25}$ at redshift 0.52;  their mass ratio,  1.23, is almost constant from z=0.42 to z=0, that is from 8.7\,Gyr.  However, the DM mass exceeds always the stellar mass within the larger radius, 50\,kpc. Their mass ratio becomes constant, about 6.2,  starting from a galaxy age of about 9\,Gyr.
This galaxy lies  in the BC for  8\,Gyr,  taking about 0.8\,Gyr to cross it and reaches the RS  at an age of 10.1\,Gyr.  
Its  SFR turns off for about 1\,Gyr  while the galaxy is crossing the GV.  From this shutdown, it arises the negative fraction of  stellar mass assembled during this phase (Table \ref{tab5}). 
Then,  the system goes back to the GV where it stays  the last 2\,Gyr until it reaches its current  position at 13.3\,Gyr.

This simulation points out that the GV is not only a transition zone  but also a region where some ETGs can come back to live for several more Gyr. 


\subsection{NGC~1533 }
\citet{Mazzei14b} already studied this galaxy,  matching some of its global properties - absolute magnitude,  SED,   {\tt Galex}  UV morphology, and kinematical properties - with a major merger (Table \ref{tab2}) between halos with perpendicular spins.
The merger occurs  at z=2.3,  that is 2.7\,Gyr after the onset of the SF.
 From the chosen snapshot, they derived  that NGC 1533 is 13.7\,Gyr old, with M$_B$=-19.9 mag.
 Here we confirm their results and  add new comparisons (see below). In particular, the SED in Fig. \ref{SEDs} includes new data from Paper II (green) and CGS catalog (black).
Comparisons of V and M2 {\tt Swift-UVOT} images (Fig. \ref{UV_OT}), as well as their luminosity profiles (Fig. \ref{prof}), are not in the previous cited paper.
The X-ray luminosity provided by the hot gas at this  age is well below  the upper limit in Table \ref{tab:tab1}.
According to this simulation, the UV ring is a resonance feature in the evolutionary scenario of this galaxy. The ring arises   when the galaxy is  about 8\,Gyr old  corresponding to z$\simeq$0.57 using  the cosmological parameters in Sect. 1.
Looking at Fig.\ref{ev_f}, the DM mass exceeds the stellar mass within 50\,kpc, by a factor that becomes almost costant at 1.26 after 8.1\,Gyr. However, within the smaller radius,  the stellar mass dominates  on the DM  starting from  5.3\,Gyr. Their mass ratio  becomes  constant ($\simeq$4)  after  8.1\,Gyr from the SF onset.
This system lies  on the BC for 7.2\,Gyr, takes about 0.9\,Gyr to cross it, and gets to the RS in 1.6\,Gyr  reaching its current position after 4\,Gyr. 
This  galaxy enters the GV at redshift higher than 0.5,  the only  case in our sample.
Table \ref{tab5} shows that   40\% of its current stellar mass is assembled before redshift 1,  and only 0.5\%, from 0.5 to 0. Moreover,  the percentage assembled moving from the brightest to the actual location in its  CMD, that is during the SF quenching, is 11\%.  

\subsection{NGC~1543}
This galaxy results from  a minor merger  (Table \ref{tab2}).
 From the snapshot which matches its  global properties,  NGC~1543  is 10.7\,Gyr old, with M$_B$=-19.7 mag.
By  averaging  the age  of its stellar populations we derive 6.5\,Gyr  and  5.6\,Gyr  weighing the same ages by B-band luminosity within a radius of 50\,kpc. 
The  SED is well matched by the selected snapshot (Fig.\ref{SEDs}). The blue data in the far-IR range are from  \citet{Temi2009}.
The shape of the FIR SED is that expected for ETGs  \citep{Mazzei94a}, however  it requires a warm/cold dust ratio of 0.05. The FIR luminosity is ~5\% of the intrinsic bolometric luminosity and 1\%  of the B luminosity ratio. 
The  amount of  cold gas  and the X-ray luminosity in Table \ref{tab:tab1} 
agree well with our results, 7.8$\times$10$^8$\,M$_{\odot}$   and   1.7$\times$10$^{38}$\,erg\,s$^{-1}$, respectively. The  average rotation velocity  of the stars, 71.4\,km\,s$^{-1}$, and their central velocity dispersion, 157\,km\,s$^{-1}$, agree well with the values in the Hyperleda catalog (70.7$\pm$14.4 and 149.4$\pm$4.4\,km\,s$^{-1}$ respectively).
Looking at its evolutionary properties (Fig. \ref{ev_f}), we note that 
the DM mass exceeds always the stellar one within the larger radius (50\,kpc). Their mass ratio becomes a constant,  $\approx$1.8, starting from an age of  about 8.4\,Gyr (z=0.18). From this point to z=0 about 4.4\% of the stellar mass is assembled  (Table \ref{tab5}).
A large percentage of the current stellar mass  (~85\%) is assembled from redshift 1, and  28\% from 0.5.  Moreover,  the percentage assembled by moving from the BP to its actual location in the CMD, that is during the quenching of the SFR, is 8\% (Table \ref{tab5}).  However,  the SFR cannot compensate for the stellar death rate  during the crossing of the GV, so that
 the stellar mass growth is negative within this radius. 
Looking at the smaller radius, R$_{25}$, the stellar mass exceeds the DM  one
starting from a galaxy age of 4.4\,Gyr, corresponding to a look-back time of 6.3\,Gyr  (z$=$0.66). Then, the stellar to DM mass ratio sets to a constant value, $\approx$3, from the entry in the RS to z=0. 
The stellar mass assembly history provides 49\% of the current  stellar mass from z=0.5 and 16\% from the BP of its CMD (Table \ref{tab5}).
While crossing   the GV the percentage of stellar mass accreted is  negative,  -2.9\%, due to the quenching of the SF for about 1\,Gyr.
However, in last 2.3\,Gyr (from z=0.18  to z=0),  the percentages of current stellar mass assembled become positive within r=50\,kpc, due to short accretion episodes  which produce galaxy rejuvenation moving back the galaxy in the GV. 

\subsection{NGC~2685}
The simulation which provides the best description of the global properties of NGC~2685 is a major merger  between two halos with perpendicular spins (Table \ref{tab2}). From these conditions arises  a late merger, occurring about 6\,Gyr from the onset of the SF.
The snapshot which best matches the global properties of this galaxy gives a galaxy age of 11.6\,Gyr and M$_B$=-19.7 mag (Table \ref{tab3p}).
By  averaging  the ages  of its stellar populations  we derive 6.2\,Gyr  and  3.2\,Gyr  weighting the  ages by B-band luminosity. 
The  total SED is well matched by the selected snapshot (Fig.\ref{SEDs}); 
(cyan) squares in the UV wavelength range are GII data from {\tt GALEX} catalogue (DR7), and   black  dots are CModel  of SDSS.
 As in the previous figures, all the data have been corrected for galactic extinction, in particular {\tt GALEX} data following \citet{Marino11a}.
The shape of the FIR SED  \citep{Mazzei94a} requires a warm to cold dust ratio of 0.1. The FIR luminosity is  about 25\%  of the intrinsic bolometric luminosity, very unusual for ETGs,  and 14 times larger than the B-band luminosity. The  amount of gas in Fig. \ref{ev_f} as well as the X-ray luminosity, $\le$ 0.01$\times$10$^{40}$\,erg\,s$^{-1}$,  are in good  agreement with  observational constraints (Table  \ref{tab:tab1}).  
The central stellar velocity dispersion (r$\le$1\,kpc) is 105\,km\,s$^{-1}$, in   agreement with the value  in  {\tt Hyperleda}   (98.9$\pm$3.6\,km\,s$^{-1}$), and the stellar rotation velocity, 110.2\,km\,s$^{-1}$,  agrees with the inclination corrected value (112.4$\pm$11.9\,km\,s$^{-1}$)  in the same catalogue.
By considering the  fixed  spherical radius of  50\,kpc, this galaxy assembles about 92\% of its current stellar mass from redshift 1, 72\% from redshift 0.5,
and only 1.6\%  while crossing the GV,   which corresponds to its current whereabouts  (Fig. \ref{ev_f} and Table \ref{tab5}).
 The corresponding fractions within R$_{25}$ (Table \ref{tab:tab1}) are 99\% , 94\%, and -0.7\%.
The DM  exceeds always the stellar mass within 50\,kpc. It becomes 2.5$\times\,$M$_{stars}$ in the past 1\,Gyr, that is z$\le$0.075,  but half  the stellar mass within R$_{25}$ in the same redshift range.
 This galaxy lies  on the BC for 9.6\,Gyr, it takes about 1.3\,Gyr to cross it, and an additional 0.7 \,Gyr  to reach its actual position on the GV.
The percentages  of current stellar mass assembled moving from the BP to the actual location in its rest-frame CMD,  that is  during the quenching of the SFR, are 16\%  and 18\%  within 50\,kpc and R$_{25}$, respectively. 

\subsection{NGC~2974}
This galaxy  is classified as E4 in {\tt RC3} 
but, as discussed in Paper~II,  with more recent classification is a SA(r)0/a. A  bright near-IR star projected to the SW  of its body,  BD-03 2751\footnote{SIMBAD database; http://simbad.u-strasbg.fr/simbad/sim-fid} complicates the picture. The JHK magnitudes in NED, by \citet{Skrutskie06}, indeed are not the total galaxy magnitudes.
We performed a new  reduction of the 2MASS images masking the star in order to derive the total galaxy near-IR magnitudes. 
Following the same method as in Paper II, we derive integrated  magnitudes  1.5 mag fainter  although with a large error:
 m$_J$=8.65 $\pm$0.48, m$_H$=7.99$\pm$0.40, m$_K$=7.78$\pm$0.49. We report these data in  Fig. \ref{SEDs}  with their 2$\sigma$ errors to describe the SED of this galaxy.
A major merger (Table \ref{tab2}) provides the best description of  its  global properties. The merger occurs 3.7\,Gyr  after the onset of the SF.
The snapshot selected provides  M$_B$=-19.9,mag and a galactic age of 12.6\,Gyr.
The  age  of the galaxy estimated by averaging ages of its stellar populations is   9.3\,Gyr and reduces to 7.8\,Gyr  
weighting their ages by B-band luminosity.
  \citet{Annibali07} found  13.9$\pm$3.61\,Gyr  for this galaxy. 
Figure \ref{SEDs}  shows the  match between the snapshot SED  (solid line) and the observed one extended over almost four orders of magnitude  in wavelength. 
The near-IR fluxes here derived are filled orange squares. 
The data  in the FIR spectral range are {\tt IRAS}  (red dots),   {\tt MIPS}  from \citet{Kaneda08}  (green), {\tt ISO} from \citet{Temi2009} (magenta),   and {\tt SCUBA} fluxes from  \citet{Savoy09}  (cyan dots). The FIR SED accounts for a  cold dust component with and  central intensity  30\% higher than that in \citet{Mazzei92}, and a warm dust component with a  temperature half that in our own Galaxy; their ratio is  r$_{w/c}$=0.3.
The FIR luminosity is 11\% of the bolometric luminosity,   one third of normal disk galaxies  \citep{Mazzei92}. 
The maximum star rotation velocity  at this snapshot, 227\,km\,sec$^{-1}$, agrees with the value in  {\tt HyperLeda}, 211.6$\pm$17.3,  and that of the gas, 110\,km\,sec$^{-1}$, with 105.2$\pm$10.4\,km\,sec$^{-1}$ in the same catalogue.
The amount of  cold gas  at the best-fit age, 1.1$\times$10$^9$\,M$\odot$ (top panel of Fig. \ref{ev_f}), is consistent with the value reported in Table \ref{tab:tab1} accounting for distance uncertainty  corresponding to a mass uncertainty of  about 20\%.
The X-ray luminosity of hot gas  at this  age, 5$\times$10$^{38}$ erg\,s$^{-1}$, is 4 times lower than the  value in Table \ref{tab:tab1}. Looking at Fig. \ref{ev_f},  the SFR  shows  a gentle self-quenching after reaching its maximum value, due to gas exhaustion and stellar feedback, lasting 2.8\,Gyr. Then, the SFR  is still on, 0.2\,M$_{\odot}/$yr  on average, unless the gas reservoir accreted.
The DM always exceeds the stellar mass  within  50\,kpc (middle right) until a stable level is achieved at an age of 7.8\,Gyr,  that is for z$\le$0.45.
The DM to stellar mass ratio is  $\approx$1.9  in this redshift range.  However, within the smaller reference radius,  R$_{25}$,  the stellar mass exceeds the DM one from redshift 0.91,  that is a galaxy age of 5\,Gyr. Their mass ratio is  $\approx 2$ for z$\le$0.5 (age$\ge$7.4\,Gyr). 
This galaxy assembles within 50\,kpc  almost 40\% of its current stellar mass at redshift higher than 1, and only $\approx$2.7\%   from 0.5, 
about 20\% and 7\% respectively, within  R$_{25}$ (Table \ref{tab5}).

\subsection{IC~2006}
We found that a major merger between two counter-rotating halos (Table \ref{tab2})  provides the best match to the global properties of this galaxy. The resulting age is 13.8\,Gyr, and M$_B$=-19.75 mag (Table \ref{tab3p}).
By  averaging  the age  of its stellar populations we derive 8.1\,Gyr  and  5.2\,Gyr  weighing the same ages by B-band luminosity. \citet{Annibali07} give 8.1$\pm$0.9 for this galaxy.
The observed SED in Fig.\ref{SEDs} (dots)  includes   all the data reported in its caption  and  IR fluxes   from {\tt SG4 GATOR} catalogue (magenta triangles). The snapshot selected provides a good match of all these data (solid line).
The  FIR SED   \citep{Mazzei94a} requires  a warm to cold dust ratio of 0.05,   and a warm dust temperature of about 55\,K.
The FIR luminosity is ~14\% of the  bolometric luminosity. 
The  amount of cold gas in Table \ref{tab:tab1} agrees with that provided by the best fit snapshot, accounting for the minimum temperature  available (8400\,K), that is 1.8$\times$10$^8$\,M$_{\odot}$. The X-ray luminosity of the hot gas is  3$\times 10^{38}$\,erg\,s$^{-1}$, somewhat lower than observed.   
The stellar central velocity dispersion is 140\,km s$^{-1}$, in agreement with 
\citet[][136$\pm$7\,km s$^{-1}$]{Kuntschner2000}. 
This galaxy lies  on the BC  12.7\,Gyr and   takes  1\,Gyr to cross the GV,  reaching its current   position after 1.1\,Gyr.
The percentages of the current stellar mass  assembled within the fixed reference radius of 50\,kpc  from redshift 1 and 0.5  correspond to $~$81\% and   52\% respectively, the corresponding fractions within R$_{25}$ are 98\% and  89\% (Table \ref{tab5}).  
The percentages  of stellar mass assembled moving from the brightest to its actual location, that is during the quenching of the SFR, are  about 10\%  and 37\% within 50\,kpc  and R$_{25}$ respectively. 
The contribution to the stellar mass growth in the GV  is only 0.7\%  within 50\,kpc, and rises to 0.9\% extending  up to z=0; it is  -2.3\%   and   -0.1\% respectively within R$_{25}$, showing that the SF turns off in the inner regions.
The DM mass exceeds always the stellar mass  within 50\,kpc. However, from an age of 11.5\,Gyr, their values  become almost the same (their mass ratio is 1.17).  
Within the smaller radius, R$_{25}$ in Table \ref{tab:tab1}, the stellar mass exceeds the DM one starting from 7.8\,Gyr , and their ratio sets to the constant value, M$_{stars}$  $\approx$4.7$\times$M$_{DM}$, from the age of about 11.5\,Gyr.
 





\end{document}